\documentclass[journal]{IEEEtran}

\usepackage[pdftex]{graphicx}
\usepackage{amsmath,bm}
\usepackage{amssymb}
\usepackage{soul,color}
\usepackage{listings}

\usepackage[T1]{fontenc}
\usepackage{hyperref}

\begin{document}
\title{Accurate modeling of continuous-time SAT solvers in SPICE}


\author{Yuriy~V.~Pershin,~\IEEEmembership{Senior~Member,~IEEE}, and Dyk Chung Nguyen,~\IEEEmembership{Student~Member,~IEEE}
\thanks{Y.~V.~Pershin and D.~C.~Nguyen are with the Department of Physics and Astronomy, University of South Carolina, Columbia, SC 29208 USA (e-mails: \mbox{pershin@physics.sc.edu} and \mbox{dykchung@email.sc.edu}).}
\thanks{Manuscript received December ..., 2024; revised ....}}

\maketitle

\begin{abstract}
Recently, there has been an increasing interest in employing dynamical systems as solvers of NP-complete problems. In this paper, we present accurate implementations of two continuous-time dynamical solvers, known in the literature as analog SAT and digital memcomputing, using advanced numerical integration algorithms of SPICE circuit simulators. For this purpose, we have developed Python scripts that convert Boolean satisfiability (SAT) problems into electronic circuits representing the analog SAT and digital memcomputing dynamical systems. Our Python scripts process conjunctive normal form (CNF) files and create netlists that can be directly imported into LTspice. We explore the SPICE implementations of analog SAT and digital memcomputing solvers by applying these to a selected set of problems and present some interesting and potentially useful findings related to digital memcomputing and analog SAT. In this work, we also introduce networks of continuous-time solvers with potential applications extending beyond the solution of Boolean satisfiability problems.
\end{abstract}


\IEEEpeerreviewmaketitle

\section{Introduction} \label{sec:1}

In recent years, there has been growing attention towards unconventional computing~\cite{Finocchio_2024}, which includes the use of physical systems for computation. In this area, the evolution of a suitable classical~\cite{SIEGELMANN1998214} or quantum~\cite{Feynman1982} system is seen as a computation process. Examples of computing with physical systems include quantum computing~\cite{bian2020solving,kowalsky20223}, memcomputing~\cite{Traversa17a,MemComputingbook}, analog SAT~\cite{zoltan}, Toshiba bifurcation machine~\cite{goto2019combinatorial}, and circuits involving stochastic tunnel junctions~\cite{borders2019integer}. In certain tasks, physical computing systems have the potential to surpass traditional von Neumann computing systems, as suggested by Arute et al.~\cite{arute2019quantum}.


 The present paper focuses on two continuous-time dynamical solvers known in the literature as analog SAT~\cite{zoltan} and digital memcomputing~\cite{Sean3SAT}. Chronologically, the first solver was the analog SAT developed by Ercsey-Ravasz and Toroczkai~\cite{zoltan}, who first proposed a mapping of $k$-SAT to a deterministic continuous-time dynamical system.
In this approach, Boolean variables are extended to continuous ones within a particular dynamical system, such that finding a solution to the $k$-SAT problem is equivalent to identifying a stable fixed point (or points) of the system. To avoid local minima, the authors of~\cite{zoltan} used a modified energy function depending on auxiliary variables. The idea is that the growth of auxiliary variables increases the weight of unresolved clauses and promotes their resolution\footnote{It is worth mentioning that the \textit{clause weighting} is also employed in certain stochastic local search algorithms~\cite{CW1,CW2,CW3,CW4,CW5}.}.
For hard instances, the analog SAT dynamics is characterized by transiently chaotic trajectories that, however, all converge to the solution~\cite{zoltan}.  It is argued that the analog SAT has a polynomial analog-time complexity~\cite{zoltan,Molnar2018}.

Digital memcomputing~\cite{Sean3SAT} is a continuous-time dynamical solver based on different equations. We emphasize that while several designs of digital memcomputing are available in the literature~\cite{Traversa17a}\footnote{See~\cite{pershin2023spice} for some crucial information to reproduce the first design~\cite{Traversa17a}.}, \cite{Sean3SAT}, here we consider the equations introduced by Bearden, Pei, and Di Ventra in~\cite{Sean3SAT} that tackle the 3-SAT problem. According to the authors, the new equations improve the analog SAT by avoiding exponential fluctuations in the energy function~\cite{Sean3SAT}.
Several properties of digital memcomputing solvers have been identified. It is argued that they rely on self-organizing logic gates to find the problem solution~\cite{Sean3SAT}. Internally, the solver operates through instantonic jumps that couple less stable critical points with more stable ones so that the number of unstable directions reduces after each jump~\cite{Bearden,Primosch23a}. Furthermore, the dynamics of digital memcomputing machines with solution(s) does not exhibit chaos and/or periodic orbits~\cite{no-chaosa,no-chaosb}.

Analog SAT and digital memcomputing dynamical systems are typically realized using numerical techniques to integrate their equations. Either conventional or dedicated hardware (such as GPUs or FPGAs) is used for this purpose.

For nearly five decades, SPICE (Simulation Program with Integrated Circuit Emphasis) circuit simulators have been in use since the publication of the original SPICE report~\cite{nagel1973simulation}. Their operation relies on sophisticated numerical integration algorithms that have been refined and tested thoroughly throughout these years. In this work, we develop Python scripts to convert the equations of analog SAT and digital memcomputing dynamical systems into equivalent electronic circuits that can be directly imported into a SPICE environment. While our work is based on LTspice XVII (Analog Device), other simulators can also be used (with suitable minor modifications).

The structure of this paper is as follows. Section~\ref{sec:2} provides an overview of the problem, solvers, and information on the SPICE modeling approach.
Examples of LTspice simulations are given in Sec.~\ref{sec:3}. These include the application of solvers to a variety of SAT problems, modification of solvers, and simulation of digital memcomputing networks.  Sec.~\ref{sec:3} is followed by a discussion in Sec.~\ref{sec:4}. The paper ends with an acknowledgment.

\renewcommand{\arraystretch}{1.5}

\begin{table*}[h]
 \caption{Summary of the analog SAT and digital memcomputing approaches}
\label{table:1}
    \begin{tabular}{ c | c | c }
\hline
     & Analog SAT~\cite{molnar2020accelerating} & Digital memcomputing~\cite{Sean3SAT}  \\ \hline
Equations   & \parbox{6cm} {\begin{eqnarray}  \dot{s}_i&=&\sum\limits_{m=1}^M2a_mc_{mi}K_{mi}(\mathbf{s})K_m(\mathbf{s}) \label{eq:aS1}\\
    \dot{a}_m&=&a_mK_m^2(\mathbf{s})\label{eq:aS2} \end{eqnarray} }  & \parbox{9.5cm} { \begin{eqnarray}
\dot{v}_n&=&\sum\limits_mx_{l,m}x_{s,m}G_{n,m}(v_n,v_j,v_k)+\left( 1+\zeta x_{l,m}\right)\cdot \nonumber \\
& & \left(1-x_{s,m}\right) R_{n,m}(v_n,v_m,v_k), \label{eq:mc1}\\
\dot{x}_{s,m}&=&\beta \left( x_{s,m} +\epsilon \right)\left( C_m(v_i,v_j,v_k)-\gamma\right) \label{eq:mc2}\\
\dot{x}_{l,m}&=&\alpha \left( C_m(v_i,v_j,v_k)-\delta\right), \label{eq:mc3} \end{eqnarray}}  \\ \hline
    Functions & \parbox{6cm} { \begin{eqnarray}  K_m&=&\frac{1}{2^3}\prod_{j=1}^N(1-c_{mj}s_j) \label{eq:aS3} \\ K_{mi}&=&\frac{K_m}{1-c_{mi}s_i} \end{eqnarray}}  &
    \parbox{9.5cm} {\begin{eqnarray}
   G_{n,m}&=&\frac{1}{2}q_{n,m}\text{min}\left[\left( 1-q_{j,m}v_j\right), \left( 1-q_{k,m}v_k\right)\right] \label{eq:mc4}\\
R_{n,m}&=&\begin{cases}
    \frac{1}{2}\left(q_{n,m}-v_n \right), \\
    \hspace{1cm} \text{if } C_m(v_n,v_j,v_k)=\frac{1}{2}\left(1-q_{n,m}v_n \right),\;\;\\
    0, \hspace{7mm} \text{otherwise}.
  \end{cases} \label{eq:mc5}\\
  C_m(v_i,v_j,v_k)&=&\frac{1}{2}\text{min}\left[\left(1-q_{i,m}v_i \right),\left(1-q_{j,m}v_j \right),\left(1-q_{k,m}v_k \right)\right]\;\;\;\;\;\;\label{eq:6}
\end{eqnarray} }
    \\ \hline
Variables &  $s_i\in[-1,1]$,\hspace{0.5cm} $i=1,..,N$      & $v_i\in[-1,1]$,\hspace{0.5cm} $i=1,..,N$         \\
         &  $a_m\in (0,\infty)$,\hspace{0.5cm} $m=1,..,M$      & $x_{s,m}\in [0,1]$, \hspace{0.5cm}$x_{l,m}\in [1,10^4M]$, \hspace{0.5cm} $m=1,..,M$              \\ \hline
Properties       & Polynomial analog-time complexity~\cite{zoltan,Molnar2018},      & Polynomial time to solution~\cite{Traversa17a,Sean3SAT}, absence of  chaos and periodic orbits in      \\
& transient chaos~\cite{zoltan} &   DMMs with solutions~\cite{Sean3SAT}, relies on instantonic dynamics~\cite{di2019digital} of SO logic gates \\ \hline
Hardware    & GPU~\cite{molnar2020accelerating}, ASIC~\cite{chang2022analog}   &  FPGA~\cite{chung23,chung24} \\
realizations        &    &  \\ \hline
    \end{tabular}
\end{table*}

\section{Methods}\label{sec:2}

\subsection{3-SAT}\label{sec:2_1}

In this paper, we limit ourselves to 3-SAT problems, wherein each clause is formed by the disjunction of 3 literals.  A literal is a Boolean variable, $x_i$, or its negation, $\bar{x}_i$.
Basically, the problem is to find the assignment of $N$ Boolean variables to satisfy $M$ clauses forming a formula. An example of a 3-SAT formula is
$$
(x_1\vee \bar{x}_2 \vee x_3)\wedge(\bar{x}_7\vee \bar{x}_5 \vee x_2)\wedge(x_3\vee \bar{x}_1 \vee x_6)\wedge...\;\;,
$$
where $\vee$ stands for the disjunction (OR) and $\wedge$ stands for the conjunction (AND). Here we just show explicitly the first three clauses.

In the above expression, the first clause can be satisfied by $x_1=\textnormal{TRUE}$, etc. However, since each variable (directly or negated) enters on average into $3M/N$ clauses, finding the solution, in general, is difficult, especially at $M/N$ close to 4.26~\cite{CRAWFORD199631}. In fact, it is well known that 3-SAT is NP-complete~\cite{lewis1983michael}.

A simple example of a 3-SAT problem can be found in the Listing~\ref{list:1} of Appendix.

\subsection{Solvers}\label{sec:2_2}

The continuous-time solvers considered in this work are summarized in Table~\ref{table:1}.

\subsubsection{Analog SAT}

The dynamical system in the analog SAT approach~\cite{molnar2020accelerating} is characterized by Eqs.~(\ref{eq:aS1}) and (\ref{eq:aS2})~\footnote{Note that in earlier publications~\cite{zoltan,Molnar2018}, Eq.~(\ref{eq:aS2}) had the form $\dot a_m=a_mK_m(s)$.}. In these equations,  $s_i$-s are continuous extensions of  Boolean variables $x_i$-s ($s_i=1$ if $x_i$ is TRUE, $s_i=-1$ if $x_i$ is FALSE), and $a_m$ are auxiliary variables. Moreover, $c_{mi}=1\;(-1)$ if $i$-th variable enters into $m$-the clause in the direct (negated) form, respectively, and $c_{mi}=0$ otherwise.

We note that Eq.~(\ref{eq:aS1}) implements the gradient descent~\cite{gu1999optimizing} for a modified energy function $V(\mathbf{s},\mathbf{a})=\sum_{m=1}^{M}a_mK_m(\mathbf{s})^2$ with $a_m$ defining the weight of $m$-th clause. According to Eq.~(\ref{eq:aS2}), the weights of unsatisfied clauses increase with time. This promotes their resolution through Eq.~(\ref{eq:aS1}).

Overall, the analog SAT is a deterministic non-local search algorithm wherein the auxiliary variables provide extra dimensions along which the trajectories escape from local minima. Several points related to Eqs.~(\ref{eq:aS1})-(\ref{eq:aS2}) should be mentioned (for more details, see~\cite{zoltan}). First, the dynamics of $\mathbf{s}$ is confined to the continuous domain $[-1,1]^N$ (for an arbitrary initial condition for $\mathbf{s}$ within $[-1,1]^N$). Second, 3-SAT solutions are stable fixed points of Eq.~(\ref{eq:aS1}) and all trajectories converge to a solution~\cite{Molnar2018}. Third, the system has no limit cycles. Fourth, for problems with solution, the only stable fixed points of Eqs.~(\ref{eq:aS1})-(\ref{eq:aS2}) are the ones corresponding to the global minimum of $V(\mathbf{s},\mathbf{a})$ with $V=0$~\cite{zoltan}.

\subsubsection{Digital memcomputing}

The dynamical system in the digital memcomputing approach~\cite{Sean3SAT} is characterized by Eqs.~(\ref{eq:mc1})-(\ref{eq:mc3}) (the original notation is used). In Eqs.~(\ref{eq:mc1})-(\ref{eq:mc3}),
$v_n$-s are continuous variables (similar to $s_i$-s in the analog SAT), $x_{s,m}$-s and $x_{l,m}$-s are the short and long memory variables, and $q_{j,m}$-s are the constants defining the clauses (same as $c_{mj}$-s in the analog SAT). The performance of digital memcomputing depends on the parameters $\alpha$, $\beta$, $\gamma$, $\delta$, $\epsilon$, and $\zeta$~\cite{Sean3SAT}.
The intervals for $v_n$, $x_{s,m}$, and $x_{l,m}$ are provided in Table~\ref{table:1}.
Moreover, Eq.~(\ref{eq:6}) defines the clause function, $C_m(v_i,v_j,v_k)$, used in Eqs.~(\ref{eq:mc2}), (\ref{eq:mc3}), and~(\ref{eq:mc5}).
This function characterizes the state of the variable that most closely satisfies clause $m$. For more information, see Refs.~\cite{Sean3SAT,MemComputingbook}.

The first term in Eq.~(\ref{eq:mc1}) can be interpreted as a ``gradient-like'' term, while the second - as a ``rigidity'' term~\cite{Sean3SAT}. The purpose of the ``rigidity'' term is to suppress the evolution of $v_n$ when its value is the best to satisfy clause $m$. Compared to a single variable $a_m$ associated with each clause in the analog SAT, in digital memcomputing, two memory variables, the short (s) and long (l), are associated with each clause.

\subsection{SPICE implementations} \label{sec:2_3}

\begin{figure}[bt]
\centering
\includegraphics[width=0.8\columnwidth]{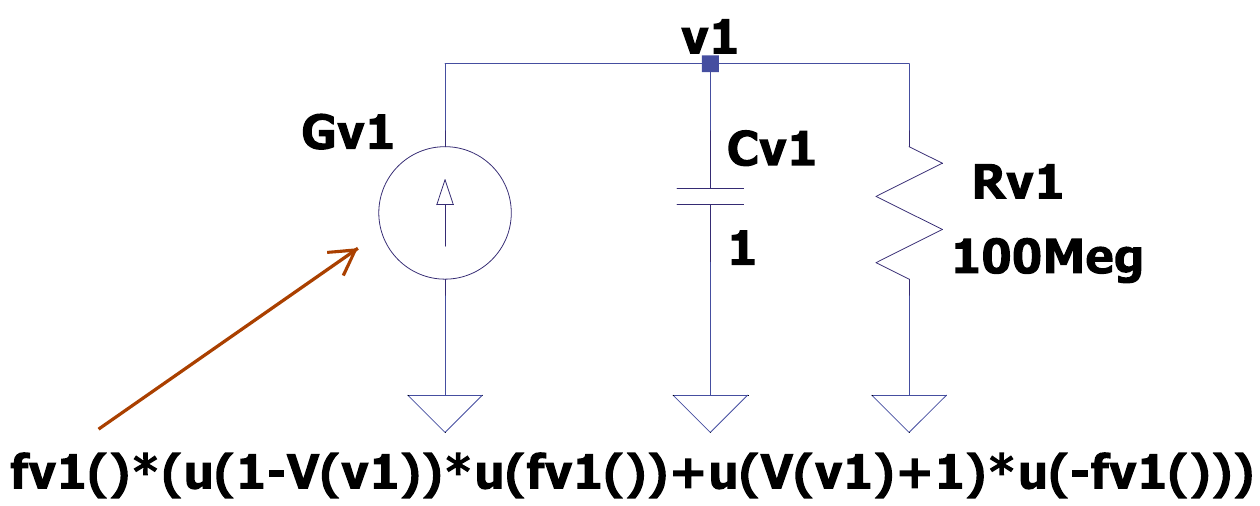}
\caption{SPICE model for the integration of Eq.~(\ref{eq:mc1}) with $n=1$. Here, $\textnormal{fv}1()$ is the function that represents the right-hand side of Eq.~(\ref{eq:mc1}). The step functions, $\textnormal{u}(\dots)$, are used to confine the variable $v_1$ to the interval $[-1,1]$.
}
\label{fig:1}
\end{figure}

Two Python scripts were developed to generate electronic circuits that implement the analog SAT and digital memory dynamical systems in SPICE. These scripts can be downloaded from~\cite{github_link_SPICE}. Each script takes a CNF file representing a 3-SAT problem as input and generates the corresponding solver circuit as output.  Python scripts are limited to clauses that contain three literals. In the case of analog SAT, the factor $1/2^3$ in Eq.~(\ref{eq:aS3}) was omitted for compactness.

In SPICE, ordinary differential equations are numerically integrated employing $1$~F capacitors driven by voltage-controlled current sources, which is a widely-used technique (see, e.g., \cite{Biolek13a}). While capacitor voltages represent variables, voltage-controlled current sources implement the right-hand sides of dynamical equations (such as Eqs.~(\ref{eq:aS1})-(\ref{eq:aS2})). As described in~\cite{Biolek13a}, to ensure a DC path to the ground, the capacitors are shunted by high-resistance resistors, which do not affect the outcome of the integration. This approach is illustrated in Fig.~\ref{fig:1}
for the variable $v_1$ in the digital memcomputing solver.

The initial conditions are set as follows. In the analog SAT solver, $s_i$-s are chosen at random from a uniform distribution in the interval $[-1,1]$ and $a_m(0)=1$. In the digital memcomputing solver, we select $v_i$-s  at random from a uniform distribution within the interval $[-1,1]$, and set $x_{s,m}(0)=0.5$ and $x_{l,m}(0)=1$. Our analog SAT and memcomputing SPICE models utilize $N+M$ and $N+2M$ capacitors, respectively, equal to the number of dynamical variables in these models.

Additionally, two control circuits are used to achieve a compact representation of the dynamics. The analog control voltage (node ``contra'') is obtained according to
\begin{equation}
V_{\textnormal{contra}}=\sum\limits_{m=1}^N C_m(v_i,v_j,v_k),
\end{equation}
where $C_m(v_i,v_j,v_k)$ is given by Eq.~(\ref{eq:6}). The digital control voltage (node ``contrd'') is derived from an analogous expression adjusted by incorporating the unit step functions, $u(\dots)$. In principle,  $V_{\textnormal{contrd}}(t)$ represents the number of unsatisfied clauses at time $t$.

Examples of SPICE netlists for the problem in the Listing~\ref{list:1} in Appendix are given in the Listings~\ref{list:2} and \ref{list:3} in Appendix.

To enable random initial values for variables, the user must make sure the "Use the clock to reseed the MC generator" option is turned on in LTSpice XVII. Moreover, the ``uic'' option should be selected in the transient analysis.

\subsection{Generation of instances}

To illustrate the approach, we used LTspice to solve easy, difficult, and very difficult 3-SAT instances with planted solutions. Easy and difficult 3-SAT instances were generated using the method of Barthel et al.~\cite{barthel2002} at $p_0=0.08$ selecting $M/N=7$ and $M/N=4.3$, respectively. For more information, see Barthel et al.~\cite{barthel2002}.

The 3 regular 3-XORSAT instances are considered challenging for various  3-SAT  solvers~\cite{kowalsky20223}. To create such an instance, a set of modulo 2 addition equations was generated via a random assignment of variables. Each equation included three variables, with their negations determined randomly. Furthermore, each variable appeared in precisely three equations. Finally, each equation was transformed into four
3-SAT clauses.

\begin{figure*}[tb]
\centering
(a)\includegraphics[width=0.45\textwidth]{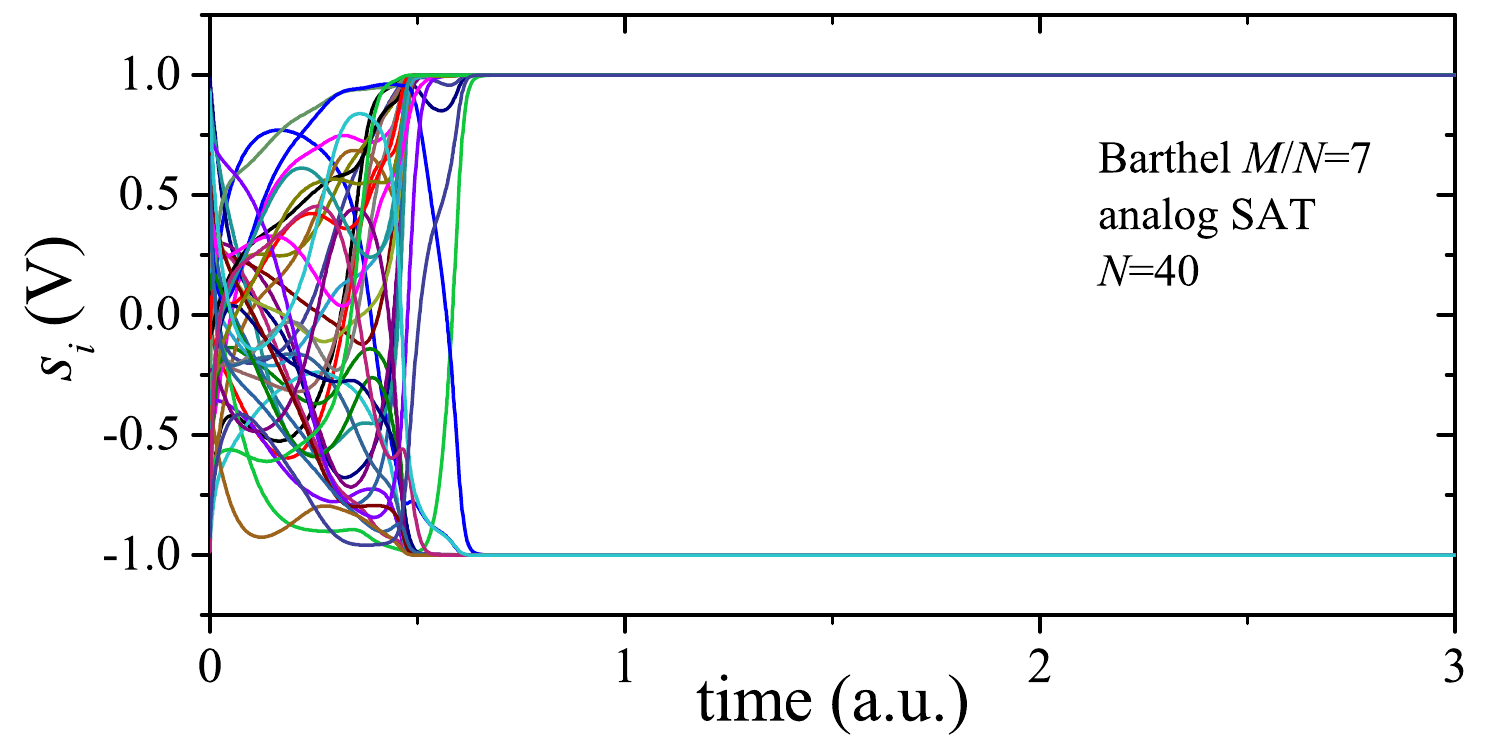}
\;(b)\includegraphics[width=0.45\textwidth]{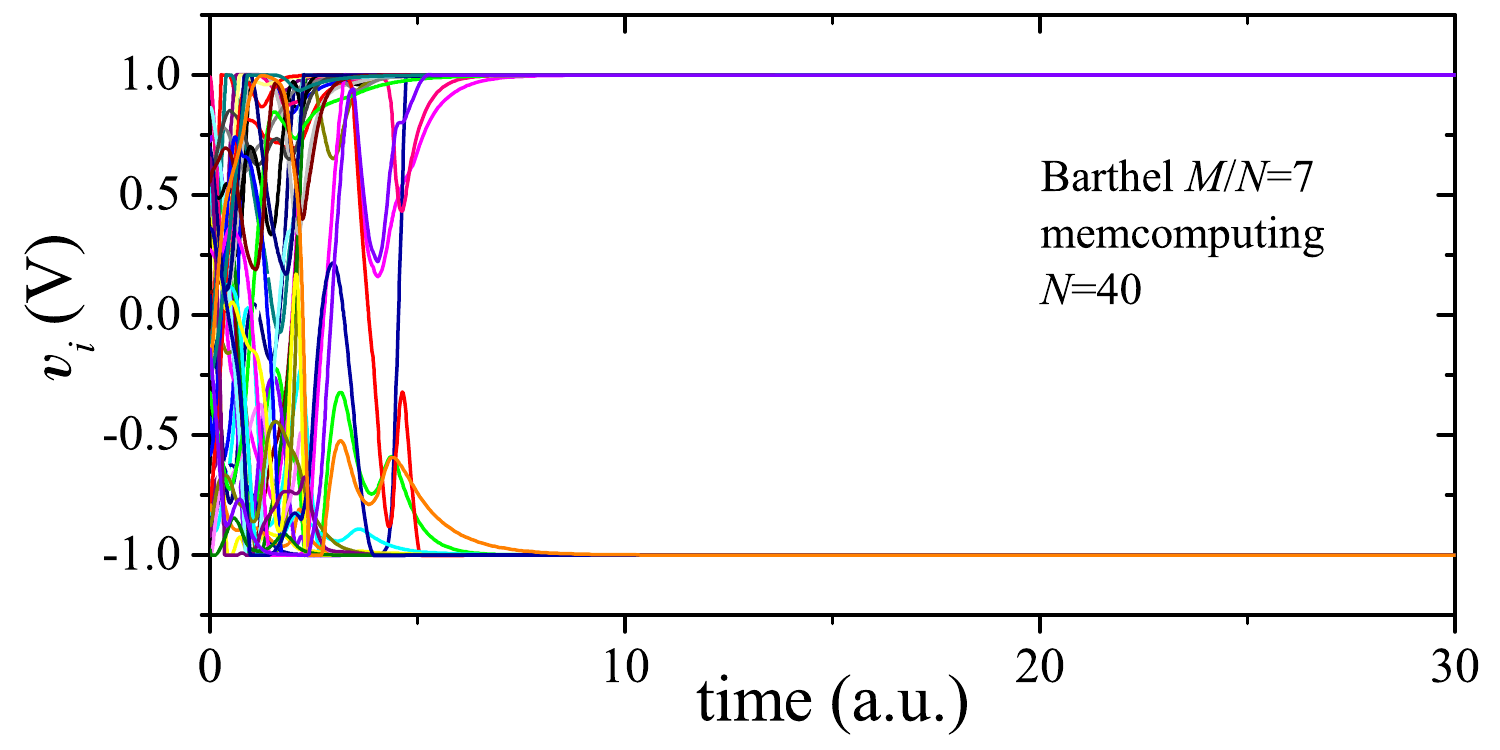} \\
(c)\includegraphics[width=0.45\textwidth]{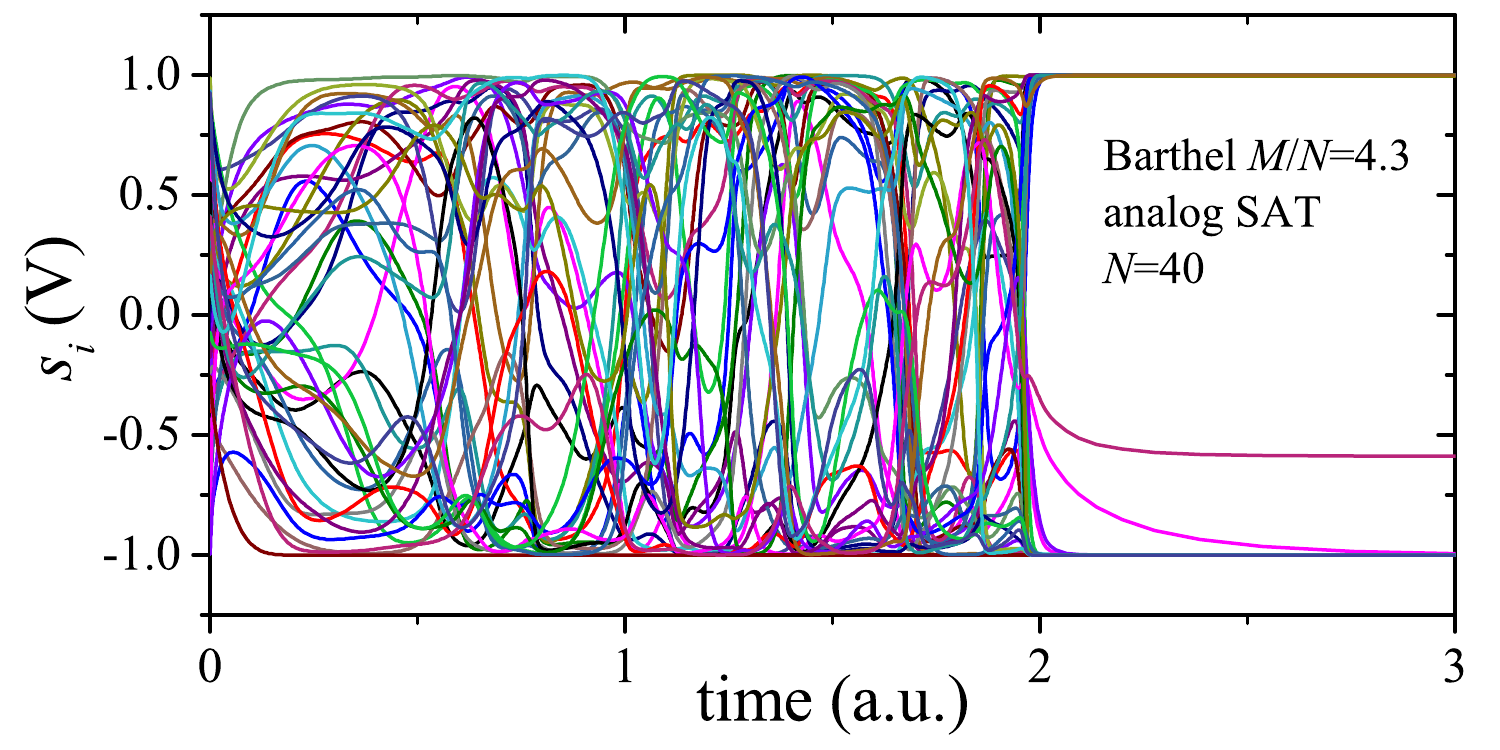}
\;(d)\includegraphics[width=0.45\textwidth]{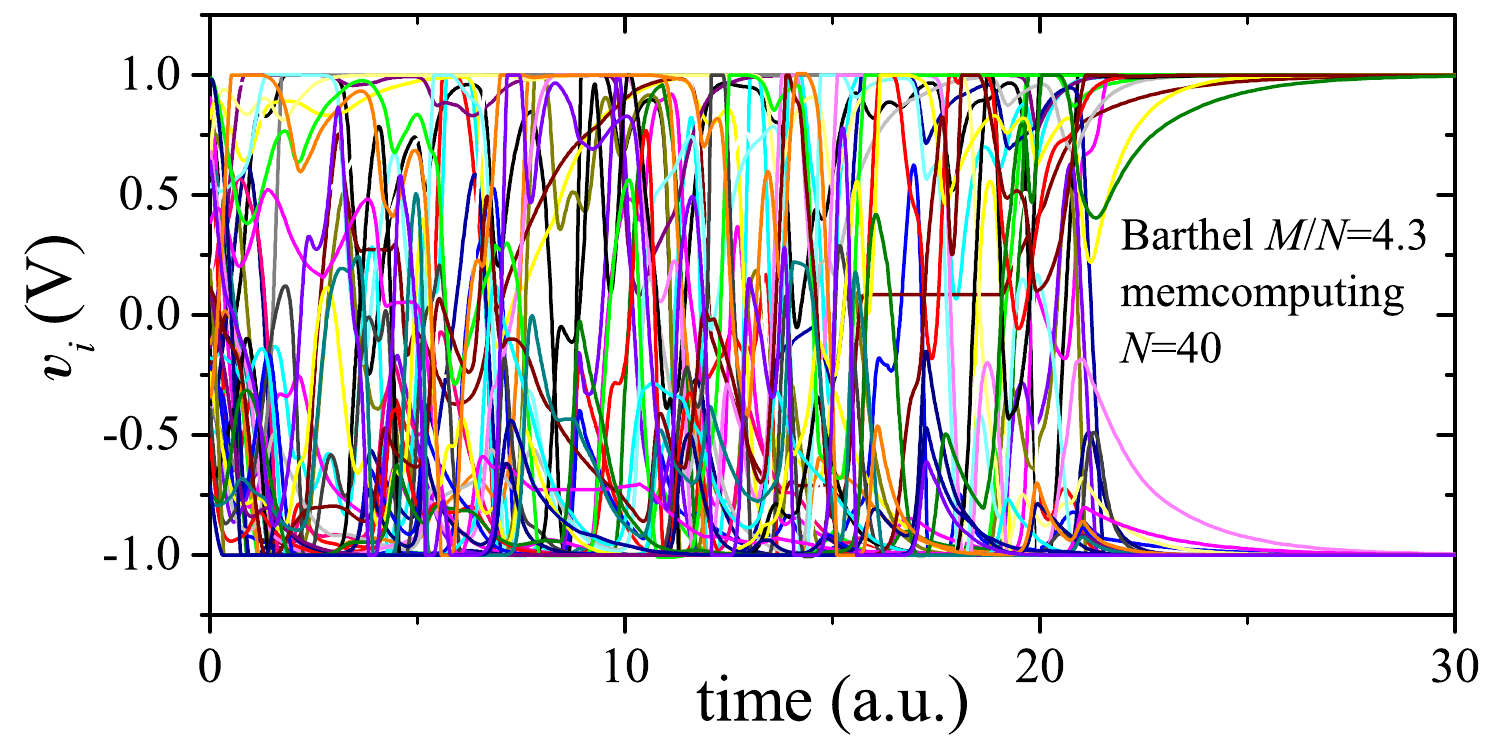}
(e)\;\includegraphics[width=0.9\textwidth]{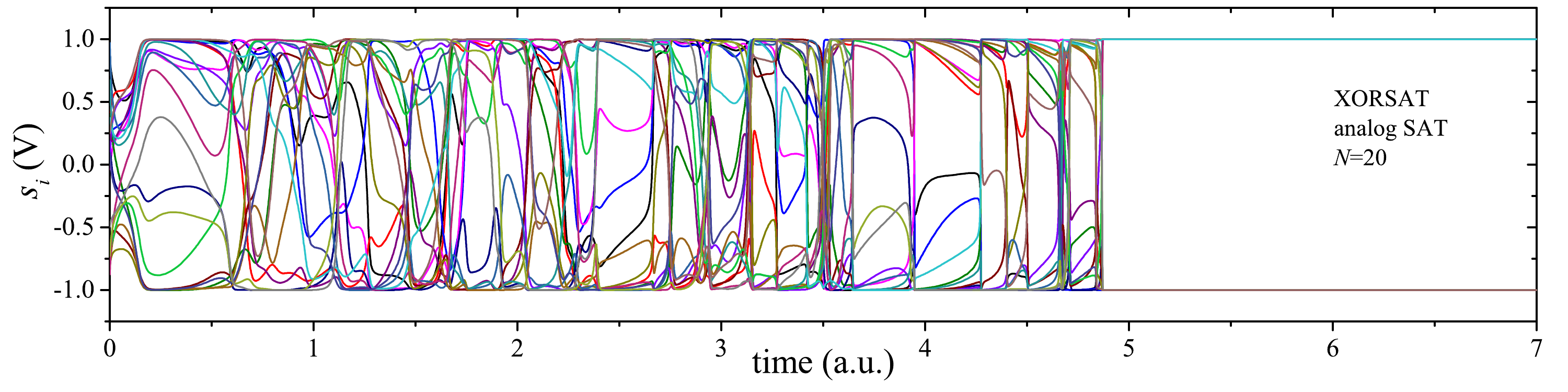} \\
(f)\;\includegraphics[width=0.9\textwidth]{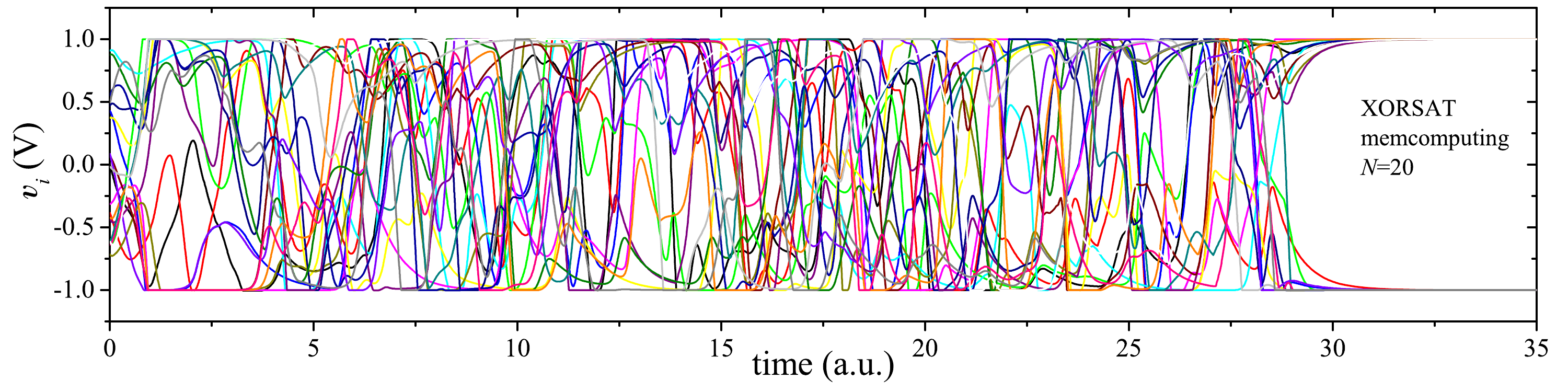}
\caption{Solving (a,b) easy, (c,d) difficult, and (e,f) very difficult 3-SAT problems in SPICE. These plots show the dynamics of the variables $s_i$ and $v_i$ in the analog SAT and digital memcomputing algorithms, respectively. These plots were obtained using  $N=40$ 3-SAT problems (panels (a)-(d)) and  one $N=20$ problem (panels (e)-(f)). Each pair of simulation results was derived from solving the same problem.
}
\label{fig:2}
\end{figure*}

\section{Simulations} \label{sec:3}

In this section, we present examples of SPICE simulations for both the analog SAT and digital memcomputing solvers. Unless specified otherwise, the digital memcomputing results were obtained using the following parameter values: $\alpha=5$, $\beta=20$, $\gamma=0.25$, $\delta=0.05$, $\epsilon=0.001$, and $\zeta=0.01$.

\subsection{Main simulations}

\begin{table}[t]
 \caption{Summary of LTspice simulations}
\label{table:2}
\centering
    \begin{tabular}{c c |  c c c}
 \hline
 & & \multicolumn{3}{c}{\# Unsolved (out of 10)} \\
  & & B4.3 & B7 & X \\
\hline
$N=10$ & analog SAT  & 0 & 0 & 10 (10) \\
  & memcomputing     & 0 & 0 & 0 \\
\hline
$N=20$ & analog SAT  & 0 & 0 & 7 (7)\\
  & memcomputing     & 0 & 0 & 0 \\
\hline
$N=30$ & analog SAT  & 0 & 0 & 8 (8)\\
  & memcomputing     & 0 & 0 & 4 \\
\hline
$N=40$ & analog SAT  & 0 & 0 & 9 (9) \\
  & memcomputing     & 0 & 0 & 10 \\
\hline
$N=50$ & analog SAT  & 0 & 0 & 10 (9)\\
  & memcomputing     & 0 & 0 &  10\\
\hline
    \end{tabular}
\end{table}

Ten instances were created for each type of the problem (easy, difficult, and very difficult) and size of the problem (we considered problems of $N=10$, $20$, $30$, $40$, and $50$). In each run, the circuit was simulated for up to $t_{ev}=300$~s of its evolution. Each solver was used on each problem just once, resulting in either a solved or unsolved outcome. For unsolved outcomes, there were two distinct scenarios: one in which the time interval $t_{ev}$ was insufficient to reach the solution and the other in which a convergence to zero was observed, as described below.

Table~\ref{table:2} provides a summary of the outcomes from our SPICE simulations. In the "Unsolved" column, the numbers in parentheses represent the number of cases when the convergence to zero was observed. The convergence to zero was only observed with the analog SAT solver.

According to Table~\ref{table:2}, all Barthel instances were solved by the two approaches. The convergence to zero  was the reason for all unsolved cases by analog SAT, except one (at $N=50$). We explain the remaining unsolved cases (by both methods) by the finite evolution time used in this study, $t_{ev}=300$~s. Presumably, this value was too short for the solvers to reach the solution. At the same time, we have not verified whether the solution can be found at large values of $t_{ev}$ in such cases.

Fig.~\ref{fig:2} presents examples of the temporal evolution of variables in solved problems of various complexity levels. According to Fig.~\ref{fig:2},  there is a clear correlation between the difficulty of the problem (easy, difficult, and very difficult) and the time to the solution. Noticeably, at any given difficulty of the problem,
it takes more time/transitions for digital memcomputing to reach the solution in comparison to the analog SAT. However, this observation is not universal, since the opposite behavior was also observed in some other cases.

\begin{figure}[h]
\centering
(a)\includegraphics[width=0.45\textwidth]{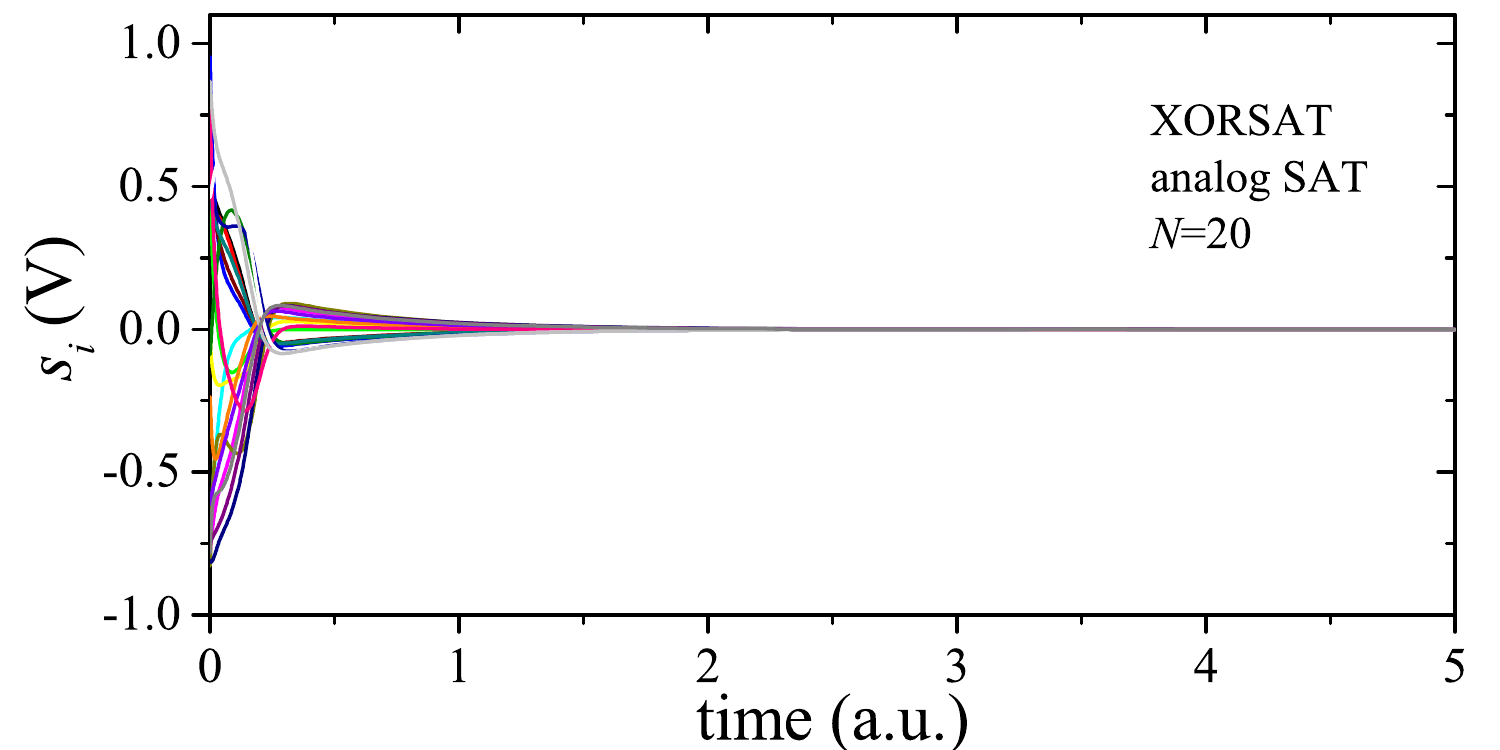}  \\
(b)\includegraphics[width=0.45\textwidth]{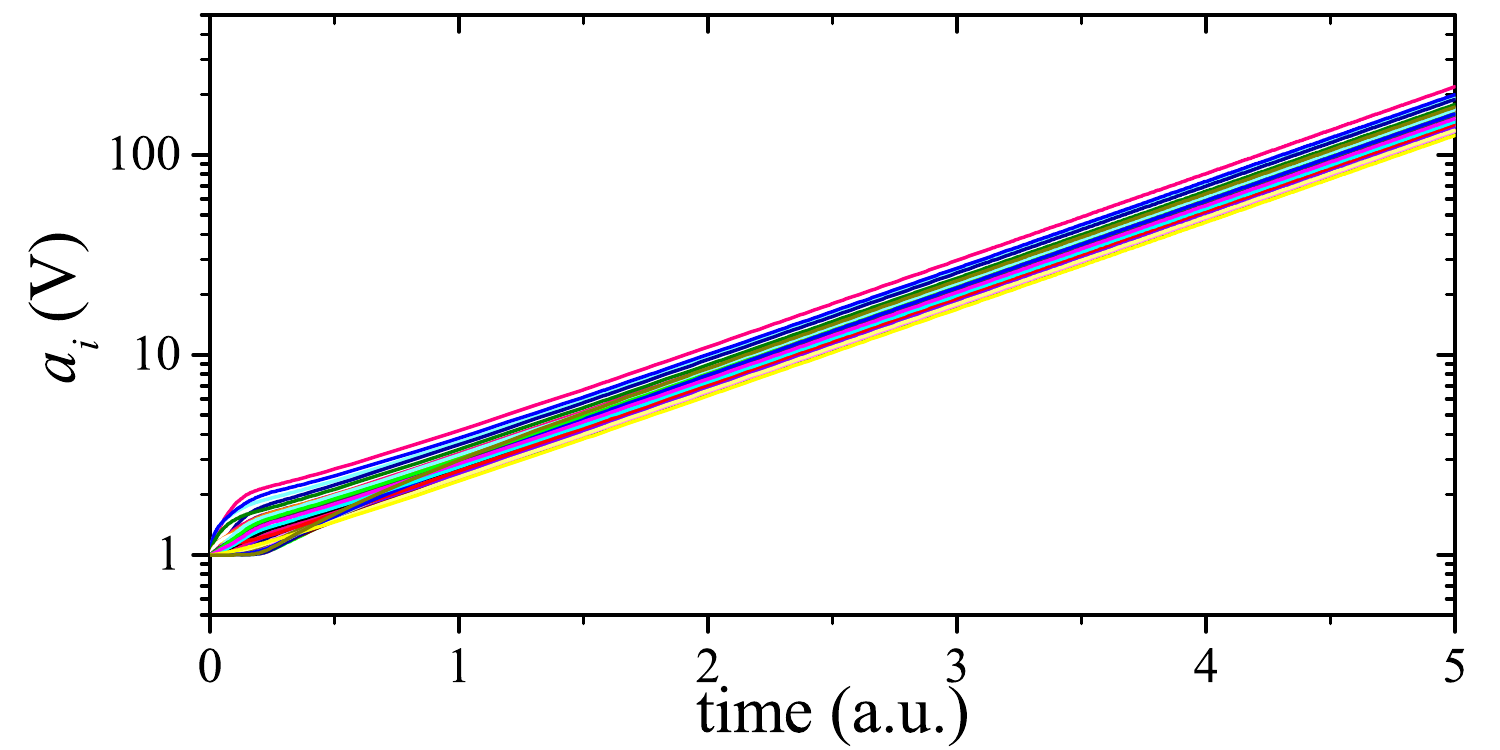}
\caption{An example of the converging to zero solution: time-dependencies of (a) main and (b) auxiliary variables. This figure was obtained using an $N=20$  3 regular 3-XORSAT problem.
}
\label{fig:3}
\end{figure}

\vspace{-0.5cm}

\subsection{Convergence to zero} \label{sec:32}

It was unexpected to observe the convergence to zero in the dynamics of analog SAT. By the ``convergence to zero'' we mean a long-term behavior when all main variables in the solver, $s_i$, approach zero as in Fig.~\ref{fig:3}(a). This transition was only observed when the analog SAT was applied to the 3 regular 3-XORSAT instances. According to the right column of Table~\ref{table:2}, the convergence to zero was the most common scenario in the case of very difficult problems.

 Figure~\ref{fig:3} illustrates how the main variables, $s_i$, and the auxiliary variables, $a_i$, behave during the convergence to zero of the main variables. It can be observed that while the main variables rapidly decrease to zero (Fig.~\ref{fig:3}(a)), the auxiliary variables exhibit exponential growth (Fig.~\ref{fig:3}(b)). It is important to point out that the convergence to zero is not the only possible scenario. It can be avoided with an alternative random selection of initial conditions.

\subsection{Modified solvers}

\begin{figure}[tb]
\centering
(a)\includegraphics[width=0.45\textwidth]{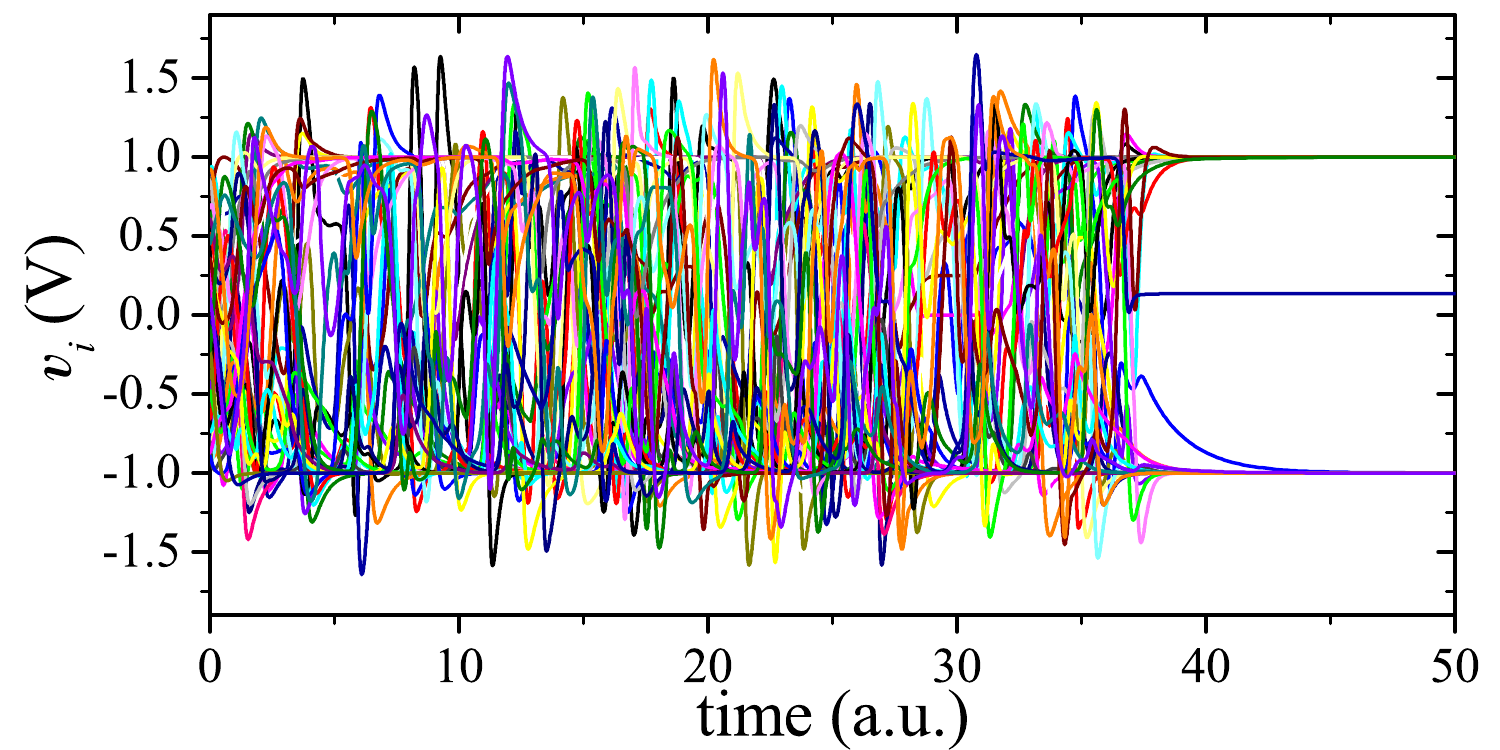} \\
(b)\includegraphics[width=0.45\textwidth]{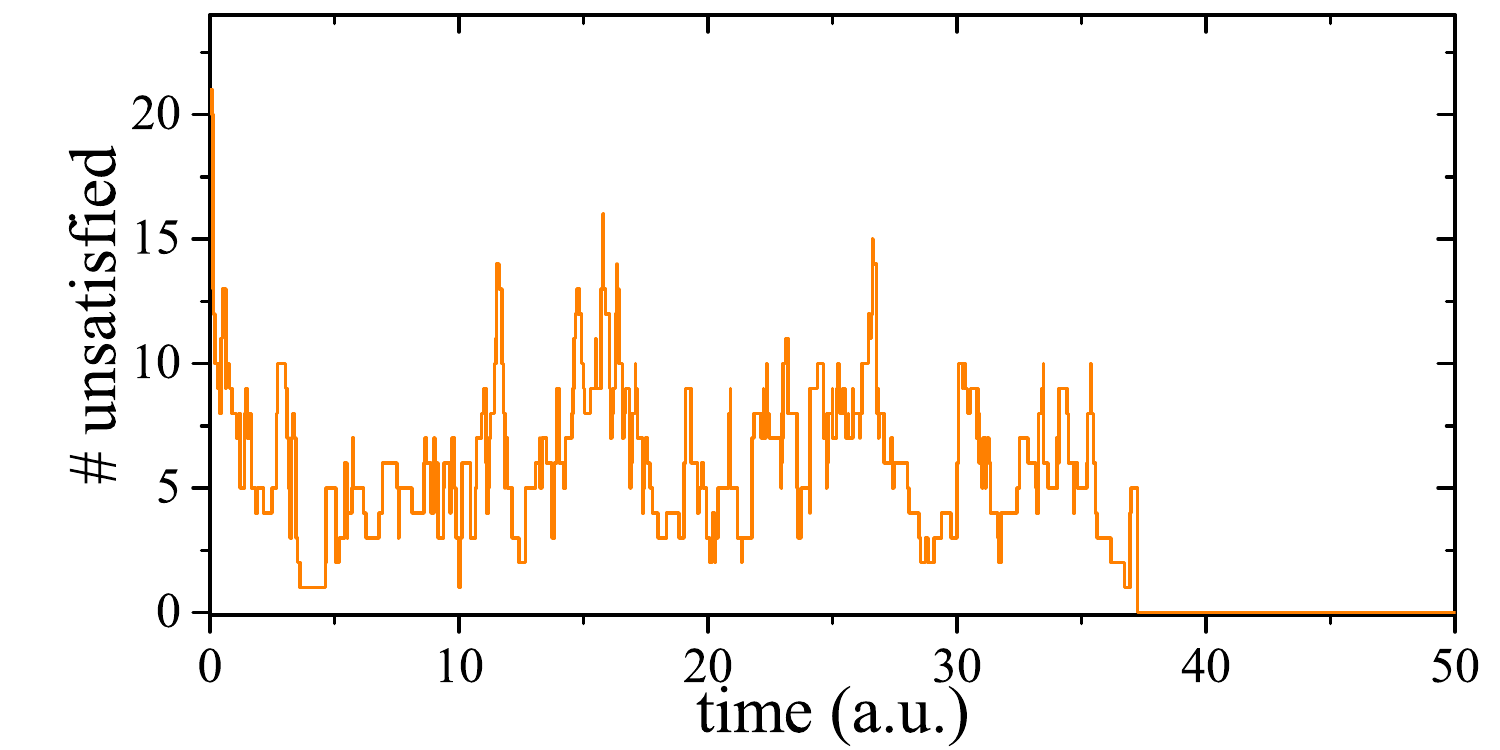} \\
\caption{Evolution of (a) voltage variables and (b) number of unsatisfied clauses in a modified memcomputing solver (without restrictions on $v_i$). The problem considered in this example is the same as in Fig.~\ref{fig:2}(c) and (d).
}
\label{fig:4}
\end{figure}

The techniques we have devised for modeling continuous-time solvers in SPICE offer a straightforward way to investigate and improve these solvers. In fact, different modifications to the solvers can be easily made, either through Python scripts or directly within the final electronic circuits. Presented below are two examples demonstrating this.

\subsubsection{Digital memcomputing}
For illustration purposes, the memcomputing solver was modified by eliminating the constraints on $v_i$ (refer to the line below Eq.~(\ref{eq:6}) in Table~\ref{table:1}). This adjustment was made directly in the solver's net list of a specific problem by removing the second multiplier from the equations that govern relevant current sources, similar to the one shown in Fig.~\ref{fig:1}. Specifically, in Fig.~\ref{fig:1} case, the expression was simplified to $\textnormal{fv1()}$.

Fig.~\ref{fig:4} shows that the digital memcomputing solver without imposing restrictions on the values $v_i$ may find the solution to the problem. Fig.~\ref{fig:4}(b) depicts the number of unsatisfied clauses as a function of time (node ``contrd''  of the circuit).

\subsubsection{Analog SAT}
In an attempt to understand the generality of the convergence to zero (Sec.~\ref{sec:32}), the following versions of Eq.~(\ref{eq:aS2}) were considered:
\begin{eqnarray}
    \dot{a}_m&=&a_mK_m(\mathbf{s})\label{eq:aS2m1}  \; ,\\
    \dot{a}_m&=&K_m(\mathbf{s})\label{eq:aS2m2} \; , \\
    \dot{a}_m&=&K_m(\mathbf{s})^2\label{eq:aS2m3} \; .
\end{eqnarray}
Using SPICE simulations, in all three cases, we have cases of the convergence to zero, as in Fig.~\ref{fig:3}(a). However, in the case of Eq.~(\ref{eq:aS2m2}) and Eq.~(\ref{eq:aS2m3}), the growth of the auxiliary variables was linear instead of exponential. Therefore, the convergence to zero seems to be unrelated to the exponential growth of the auxiliary variables shown in Fig.~\ref{fig:3}(b).

\subsection{Networks of SAT solvers} \label{sec:D}

\begin{figure*}[tb]
\centering
(a)\includegraphics[width=0.45\textwidth]{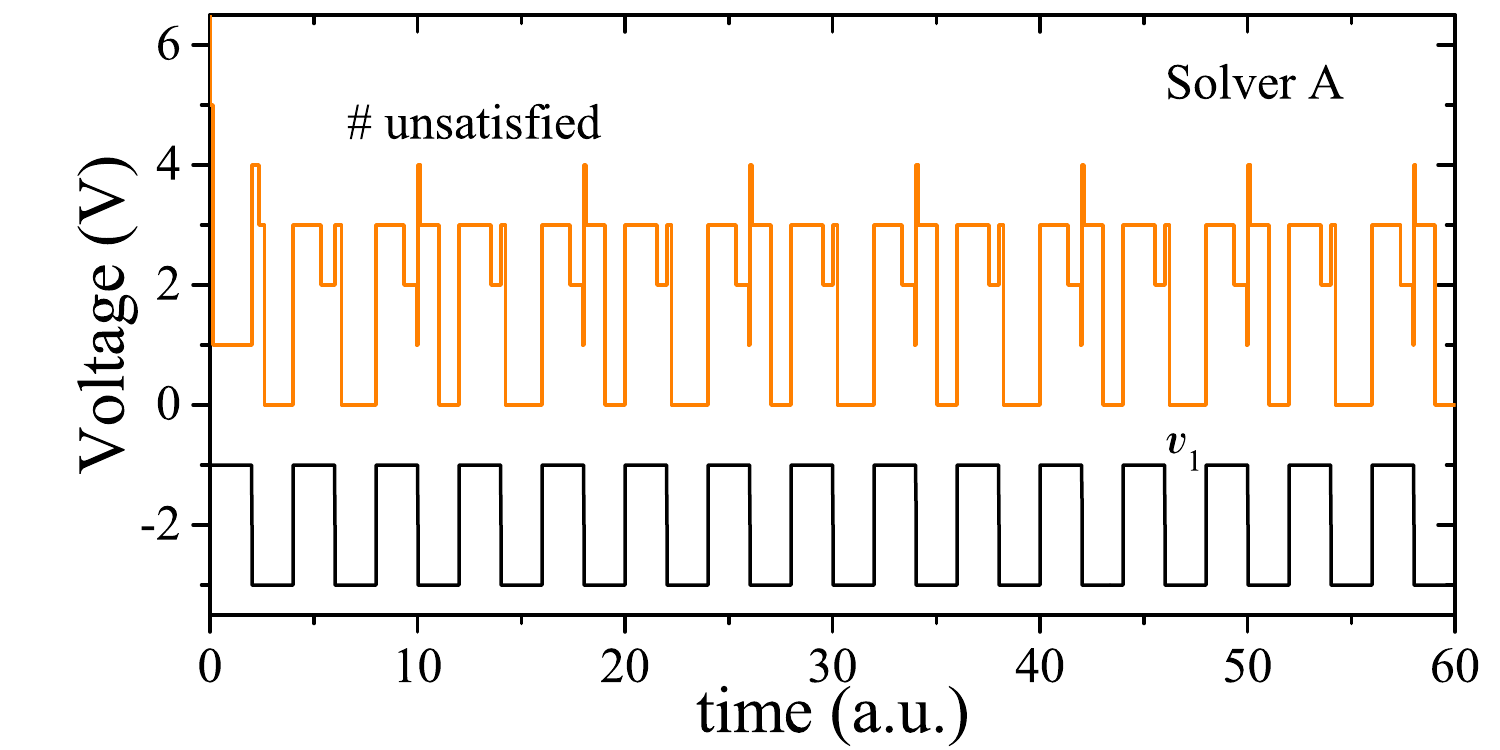}  \;
(c)\includegraphics[width=0.45\textwidth]
{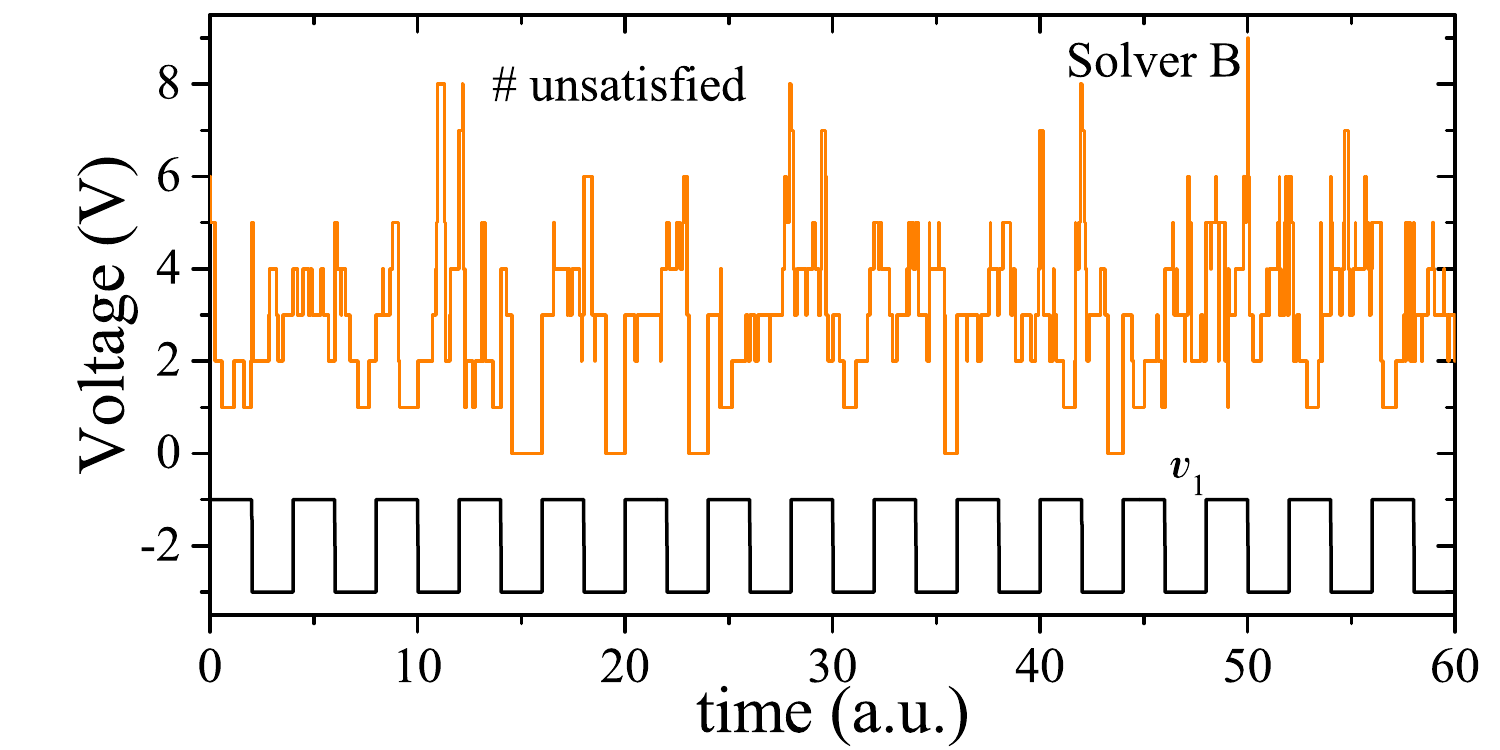}  \\
(b)\includegraphics[width=0.45\textwidth]
{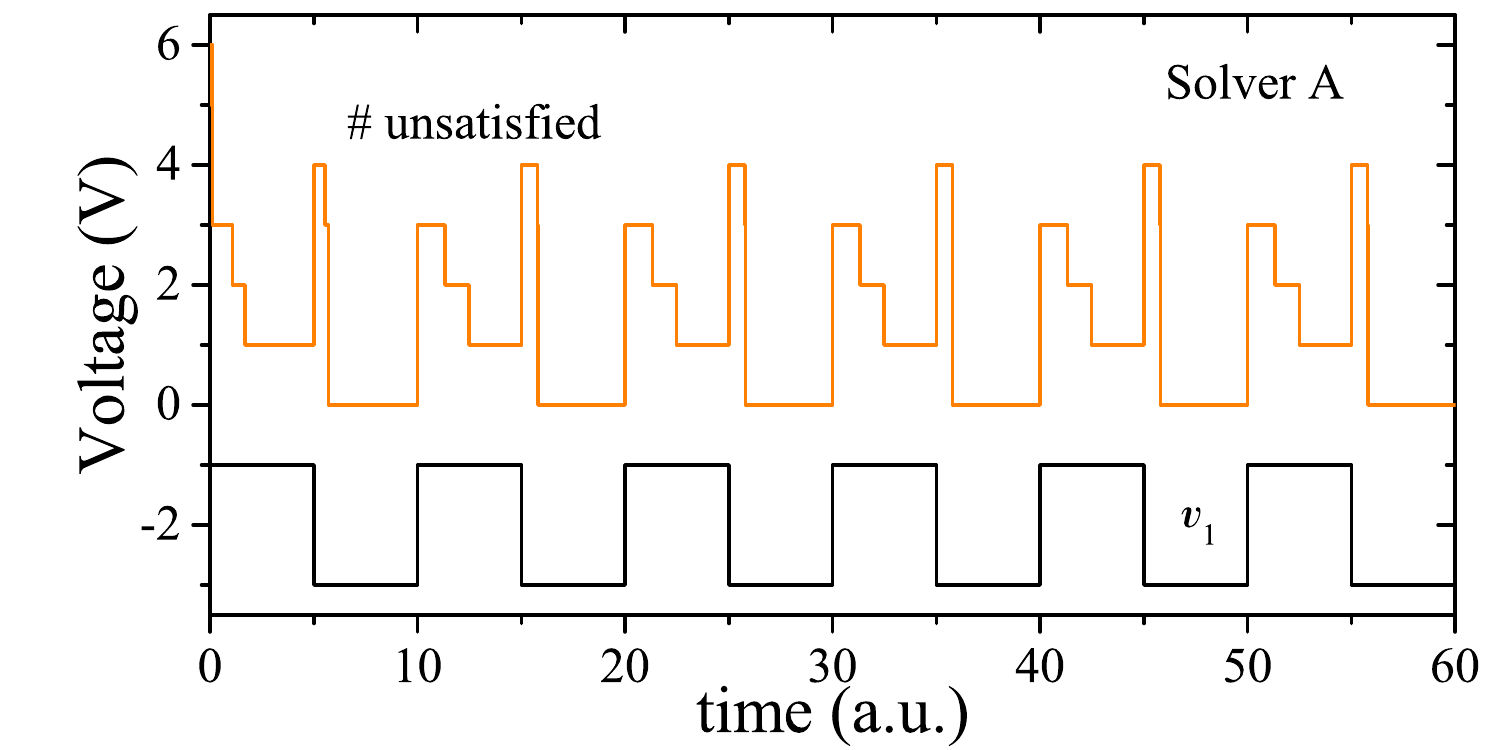} \;
(d)\includegraphics[width=0.45\textwidth]{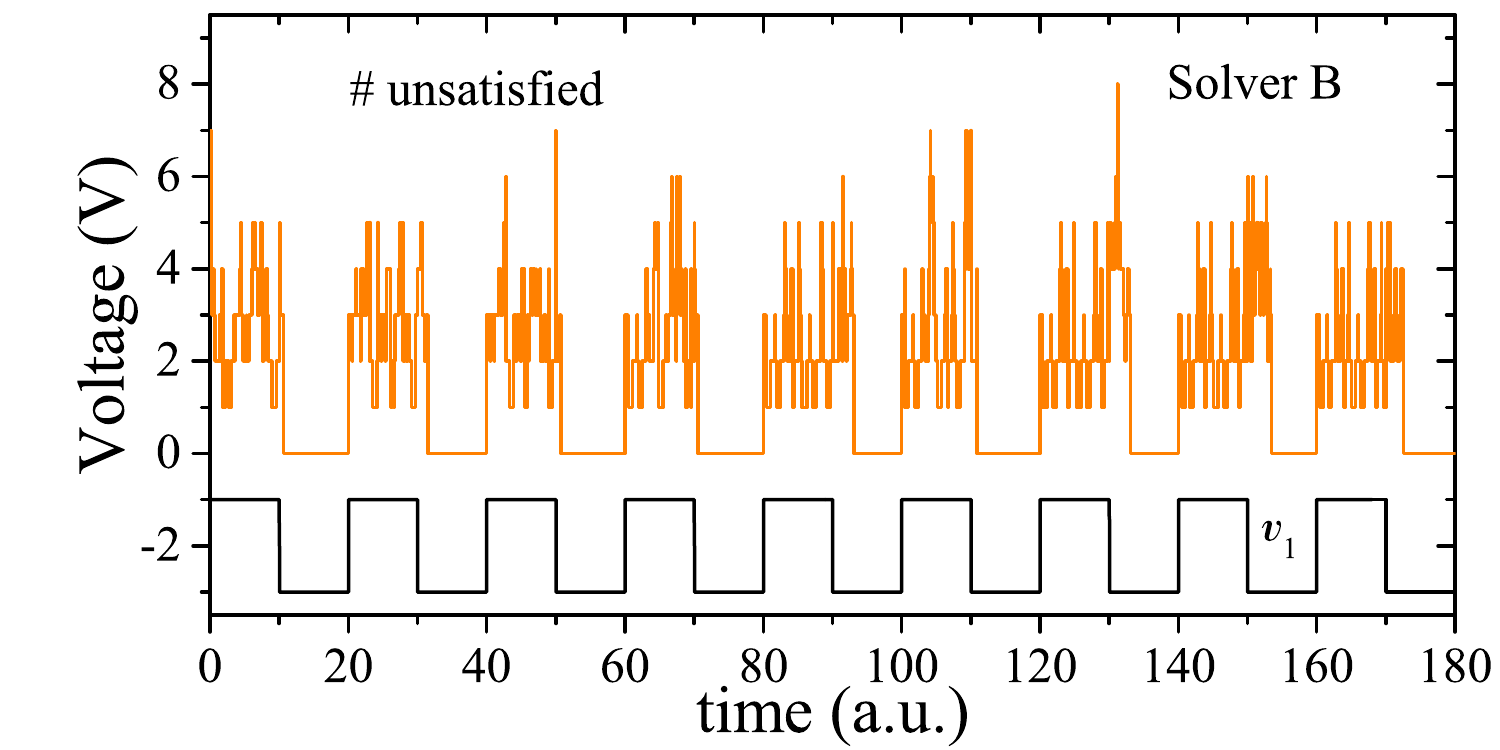} \\
\caption{Solvers A and B's response to a square wave voltage (represented by $v_1$). For the sake of clarity, the applied voltage curves were shifted downward by $2$~V. See Sec.~\ref{sec:D} for more details. These simulations were performed using $\alpha=0$.
}
\label{fig:5}
\end{figure*}

It might be worthwhile to explore networks composed of continuous-time SAT solvers, where these solvers serve as nodes that interact via some of their variables or combinations of them. On the one hand, SAT solver networks can be utilized to solve SAT problems whose size exceeds the capacity of an individual SAT solver, a particularly crucial approach for physical implementations of these solvers that have finite sizes~\cite{chang2022analog}. On the other hand, SAT solvers can be viewed as distant analogs to neurons, possessing complex functionalities determined by the SAT problems they solve. Therefore, one can anticipate that networks of SAT solvers could find applications in artificial neural network architectures or reservoir computing.

An especially useful feature of SPICE for this application is the subcircuit declaration. Fundamentally, the subcircuit functions as a collection of elements in SPICE that can be referred to in a manner similar to device models. Externally, the subcircuit is accessed through its external nodes. For SAT solvers, a potential approach is to define $P$ input nodes and $Q$ output nodes ($P+Q\leq N$), such that, in the digital memcomputing case, the input nodes simply set the values of $P$ voltage variables (their dynamics equations are disregarded), while the output nodes are used to output the values of some other $Q$ voltage variables. To monitor the solver's state, the signal $V_{\textnormal{contrd}}(t)$ may be included in the output nodes. Of course, there are other choices available to connect the solvers.

\begin{figure*}[h]
\centering
(a) \hspace{0.7cm} \includegraphics[width=0.3\textwidth]{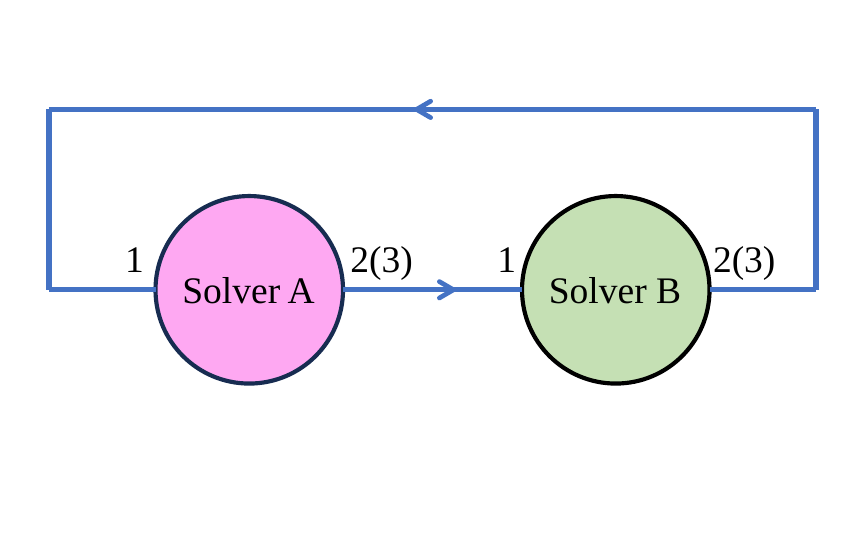}
\hspace{1cm}
(c) \includegraphics[width=0.45\textwidth]{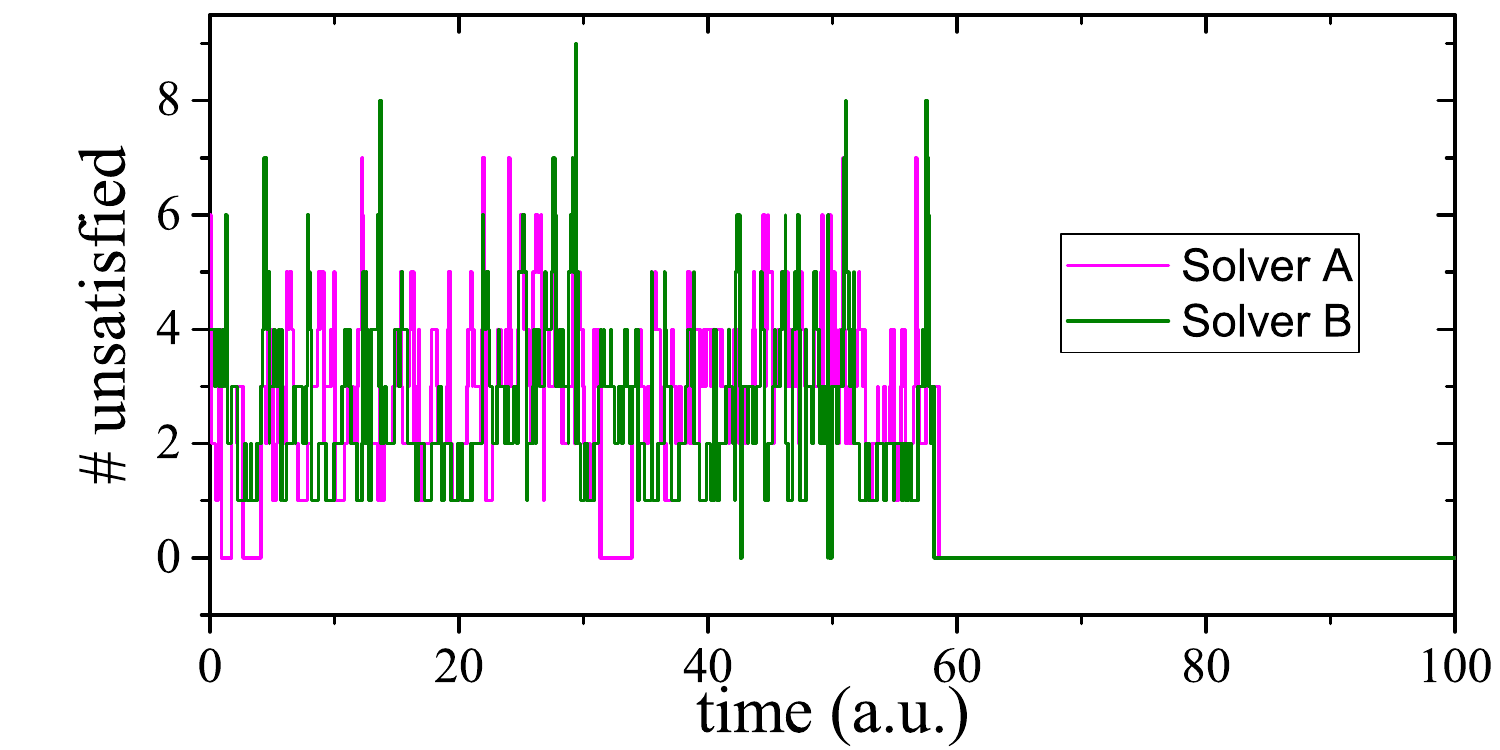}  \\
\vspace{2mm}
(b) \includegraphics[width=0.9\textwidth]{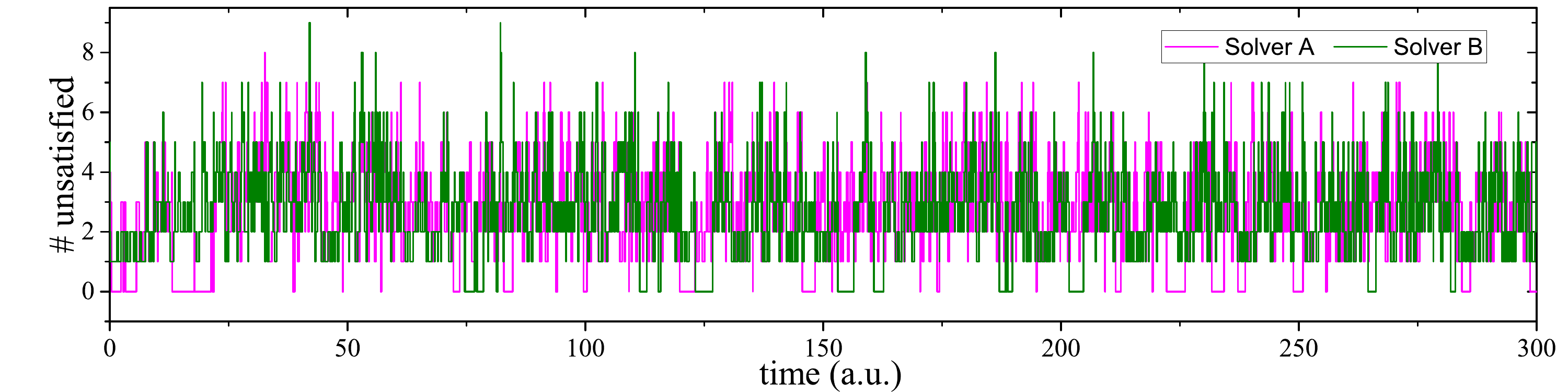}
\caption{ (a) Schematics of a simple network of two digital memcomputing solvers. (b) Number of unsatisfied clauses as a function of time when $v_2$ is used as the output node in both solvers. (c) Number of unsatisfied clauses as a function of time when $v_3$ is used as the output node in both solvers.
}
\label{fig:6}
\end{figure*}

As an example, we considered two instances, say A and B, of the 3-regular 3-XORSAT problem with $N=10$. The digital memcomputing circuits for A and B were created using our Python script~\cite{github_link_SPICE}. First, we simulated uncoupled solvers and found the following solutions (which are not necessarily unique):
\begin{eqnarray}
    && 0101100111 \;\;\;\;\;\;\;\;\; \textnormal{for instance A,} \label{eq:solA}
    \\
    && 0100111101 \;\;\;\;\;\;\;\;\; \textnormal{for instance B.}\label{eq:solB}
\end{eqnarray}
Here, each digit from left to right represents $x_1$ through $x_{10}$.
By modifying several lines in the netlists of solvers A and B, we  created the subcircuits for subsequent simulations. The sets $\{ v_1\}$, $\{ v_i, V_{\textnormal{contrd}}(t)\}$ with $i=2$ or $3$ were selected as input and output nodes, respectively.

Next, we studied how individual solvers react to an applied square wave voltage. To achieve this, a square wave voltage source ($-1$~V to $+1$~V) was connected to the solver's input node. We have observed that when the applied voltage is at logic zero and there is sufficient time, both solvers can successfully identify the solution. When the applied voltage is at logic one, the solvers exhibit some interesting dynamics, which can be described as deterministic for solver A and chaotic for solver B. These observations are evident in Fig.~\ref{fig:5}. A notable characteristic associated with solver A is the period doubling behavior depicted in Fig.~\ref{fig:5}(a), which vanishes at a longer period, as demonstrated in Fig.~\ref{fig:5}(b).

Finally, the solvers were connected in a network as shown in Fig.~\ref{fig:6}(a). Two cases were considered: in the first case,  $v_2$ was used as the output node in both solvers, while in the second case, $v_3$ was employed for the same purpose. The networks were simulated for 300 time units of the circuit dynamics. Fig.~\ref{fig:6}(b) demonstrates that selecting $v_2$ as the output node results in no solution for the whole network, whereas using $v_3$ as the output node successfully yields a solution. These observations align with the solutions provided by Eqs.~(\ref{eq:solA}) and (\ref{eq:solB}) and can be explained by them.

\section{Discussion}\label{sec:4}

In this work, we have developed a platform for high-accuracy modeling of continuous-time SAT solvers within the SPICE framework. To achieve this, we utilized Python scripts to produce electronic circuits that implement continuous-time solvers of specific problems. Cutting-edge SPICE simulation tools, such as LTspice, used in this research are based on advanced numerical algorithms to simulate circuit dynamics. These algorithms being a part of our platform are used to solve NP-complete problems via circuit simulations. On a standard PC, LTspice can effectively handle problems with $N\lesssim 50$. For larger 3-SAT problems, massively parallel SPICE circuit simulators such as Xyse$^{\textnormal{TM}}$~\cite{doecode_2462} could offer potential solutions.

Through the application of analog SAT and digital memcomputing techniques to 3-SAT problems of several levels of complexity, we have observed, for the first time,  the transition to zero in the behavior of certain analog SAT dynamical systems. Such behavior is robust with respect to the equations that define the dynamics of auxiliary variables and have a large basin of attraction.
This observation challenges the statement that ``eventually all trajectories converge to a solution''~\cite{Molnar2018}.

Additionally, we have observed that the digital memcomputing solver continues to function effectively when the limitation on voltage variables is removed. This observation could prove beneficial in the design of specialized digital memcomputing hardware~\cite{zhang2024implementation}.

Lastly, we have demonstrated that it is quite straightforward to simulate networks of continuous-time solvers in SPICE.
It is expected that such networks may have potential applications extending beyond the solution of Boolean satisfiability problems.

In a wider perspective, although digital memcomputing and analog SAT belong to the same group of algorithms, their development has occurred largely in isolation. This presents opportunities for mutual enrichment of these methodologies at both theoretical and applied levels. For example, digital memcomputing operations are believed to incorporate instantonic jumps that connect less stable critical points with more stable ones, thereby decreasing the number of unstable directions after each jump~\cite{Primosch23a}. We propose that analogous instantonic jumps may be present in analog SAT as well. In fact, in Figs.~\ref{fig:2}(c) and (e) showing the dynamics of analogue SAT, it is easy to identify distinct groups of transitions resembling those previously attributed to instantons in digital memcomputing~\cite{Primosch23a}.
A similar argument applies to the role of self-organizing gates in these solvers.


\section*{Acknowledgements}
The authors acknowledge the support of the National
Science Foundation grant number ECCS-2229880. They thank M. Di Ventra and Y.-H. Zhang for comments on the manuscript.



\lstset{basicstyle=\ttfamily\footnotesize,breaklines=true}

\newpage

\appendices

\section{File listings} \label{app:1}
\begin{lstlisting}[caption={A simple 3-SAT problem in CNF notation}, label=list:1]
c p0=0.08
p cnf 5 15
5 -4 3 0
5 -2 -3 0
3 5 1 0
-2 5 3 0
-2 1 -5 0
-4 -1 3 0
5 -1 -4 0
5 3 4 0
-5 2 -4 0
-3 -5 -2 0
-3 5 4 0
3 -4 2 0
-2 4 5 0
1 3 5 0
1 3 4 0
\end{lstlisting}

\begin{lstlisting}[caption={LTspice netlist  for the analog SAT algorithm for the problem in Listing~\ref{list:1}}, label=list:2]
* Control circuit
ESAT1 contra 0 value={fsat1()}
RSAT1 contra 0 100meg
ESAT2 contrd 0 value={fsat2()}
RSAT2 contrd 0 100meg

* Main variables
Cs1 s1 0 1 IC={-1+mc(1,1)}
Gs1 0 s1 value={fs1()}
Rs1 s1 0 100meg
Cs2 s2 0 1 IC={-1+mc(1,1)}
Gs2 0 s2 value={fs2()}
Rs2 s2 0 100meg
Cs3 s3 0 1 IC={-1+mc(1,1)}
Gs3 0 s3 value={fs3()}
Rs3 s3 0 100meg
Cs4 s4 0 1 IC={-1+mc(1,1)}
Gs4 0 s4 value={fs4()}
Rs4 s4 0 100meg
Cs5 s5 0 1 IC={-1+mc(1,1)}
Gs5 0 s5 value={fs5()}
Rs5 s5 0 100meg

* Memory variables
Ca1 a1 0 1 IC={1}
Ga1 0 a1 value={fa1()}
Ra1 a1 0 100meg
Ca2 a2 0 1 IC={1}
Ga2 0 a2 value={fa2()}
Ra2 a2 0 100meg
Ca3 a3 0 1 IC={1}
Ga3 0 a3 value={fa3()}
Ra3 a3 0 100meg
Ca4 a4 0 1 IC={1}
Ga4 0 a4 value={fa4()}
Ra4 a4 0 100meg
Ca5 a5 0 1 IC={1}
Ga5 0 a5 value={fa5()}
Ra5 a5 0 100meg
Ca6 a6 0 1 IC={1}
Ga6 0 a6 value={fa6()}
Ra6 a6 0 100meg
Ca7 a7 0 1 IC={1}
Ga7 0 a7 value={fa7()}
Ra7 a7 0 100meg
Ca8 a8 0 1 IC={1}
Ga8 0 a8 value={fa8()}
Ra8 a8 0 100meg
Ca9 a9 0 1 IC={1}
Ga9 0 a9 value={fa9()}
Ra9 a9 0 100meg
Ca10 a10 0 1 IC={1}
Ga10 0 a10 value={fa10()}
Ra10 a10 0 100meg
Ca11 a11 0 1 IC={1}
Ga11 0 a11 value={fa11()}
Ra11 a11 0 100meg
Ca12 a12 0 1 IC={1}
Ga12 0 a12 value={fa12()}
Ra12 a12 0 100meg
Ca13 a13 0 1 IC={1}
Ga13 0 a13 value={fa13()}
Ra13 a13 0 100meg
Ca14 a14 0 1 IC={1}
Ga14 0 a14 value={fa14()}
Ra14 a14 0 100meg
Ca15 a15 0 1 IC={1}
Ga15 0 a15 value={fa15()}
Ra15 a15 0 100meg

* functions
.func Cm(x,y,z)={0.5*min(1-x,min(1-y,1-z))}
.func Cm1(x,y,z)={min(1-u(x),min(1-u(y),1-u(z)))}

.func fsat1()=Cm(V(s5),-V(s4),V(s3))+Cm(V(s5),-V(s2),-V(s3))+Cm(V(s3),V(s5),V(s1))+Cm(-V(s2),V(s5),V(s3))+Cm(-V(s2),V(s1),-V(s5))+Cm(-V(s4),-V(s1),V(s3))+Cm(V(s5),-V(s1),-V(s4))+Cm(V(s5),V(s3),V(s4))+Cm(-V(s5),V(s2),-V(s4))+Cm(-V(s3),-V(s5),-V(s2))+Cm(-V(s3),V(s5),V(s4))+Cm(V(s3),-V(s4),V(s2))+Cm(-V(s2),V(s4),V(s5))+Cm(V(s1),V(s3),V(s5))+Cm(V(s1),V(s3),V(s4))

.func fsat2()=Cm1(V(s5),-V(s4),V(s3))+Cm1(V(s5),-V(s2),-V(s3))+Cm1(V(s3),V(s5),V(s1))+Cm1(-V(s2),V(s5),V(s3))+Cm1(-V(s2),V(s1),-V(s5))+Cm1(-V(s4),-V(s1),V(s3))+Cm1(V(s5),-V(s1),-V(s4))+Cm1(V(s5),V(s3),V(s4))+Cm1(-V(s5),V(s2),-V(s4))+Cm1(-V(s3),-V(s5),-V(s2))+Cm1(-V(s3),V(s5),V(s4))+Cm1(V(s3),-V(s4),V(s2))+Cm1(-V(s2),V(s4),V(s5))+Cm1(V(s1),V(s3),V(s5))+Cm1(V(s1),V(s3),V(s4))

.func fs1() = {2*V(a3)*(1)*pow(1-V(s3),2)*pow(1-V(s5),2)*pow(1-V(s1),1)+2*V(a5)*(1)*pow(1+V(s2),2)*pow(1-V(s1),1)*pow(1+V(s5),2)+2*V(a6)*(-1)*pow(1+V(s4),2)*pow(1+V(s1),1)*pow(1-V(s3),2)+2*V(a7)*(-1)*pow(1-V(s5),2)*pow(1+V(s1),1)*pow(1+V(s4),2)+2*V(a14)*(1)*pow(1-V(s1),1)*pow(1-V(s3),2)*pow(1-V(s5),2)+2*V(a15)*(1)*pow(1-V(s1),1)*pow(1-V(s3),2)*pow(1-V(s4),2)}
.func fs2() = {2*V(a2)*(-1)*pow(1-V(s5),2)*pow(1+V(s2),1)*pow(1+V(s3),2)+2*V(a4)*(-1)*pow(1+V(s2),1)*pow(1-V(s5),2)*pow(1-V(s3),2)+2*V(a5)*(-1)*pow(1+V(s2),1)*pow(1-V(s1),2)*pow(1+V(s5),2)+2*V(a9)*(1)*pow(1+V(s5),2)*pow(1-V(s2),1)*pow(1+V(s4),2)+2*V(a10)*(-1)*pow(1+V(s3),2)*pow(1+V(s5),2)*pow(1+V(s2),1)+2*V(a12)*(1)*pow(1-V(s3),2)*pow(1+V(s4),2)*pow(1-V(s2),1)+2*V(a13)*(-1)*pow(1+V(s2),1)*pow(1-V(s4),2)*pow(1-V(s5),2)}
.func fs3() = {2*V(a1)*(1)*pow(1-V(s5),2)*pow(1+V(s4),2)*pow(1-V(s3),1)+2*V(a2)*(-1)*pow(1-V(s5),2)*pow(1+V(s2),2)*pow(1+V(s3),1)+2*V(a3)*(1)*pow(1-V(s3),1)*pow(1-V(s5),2)*pow(1-V(s1),2)+2*V(a4)*(1)*pow(1+V(s2),2)*pow(1-V(s5),2)*pow(1-V(s3),1)+2*V(a6)*(1)*pow(1+V(s4),2)*pow(1+V(s1),2)*pow(1-V(s3),1)+2*V(a8)*(1)*pow(1-V(s5),2)*pow(1-V(s3),1)*pow(1-V(s4),2)+2*V(a10)*(-1)*pow(1+V(s3),1)*pow(1+V(s5),2)*pow(1+V(s2),2)+2*V(a11)*(-1)*pow(1+V(s3),1)*pow(1-V(s5),2)*pow(1-V(s4),2)+2*V(a12)*(1)*pow(1-V(s3),1)*pow(1+V(s4),2)*pow(1-V(s2),2)+2*V(a14)*(1)*pow(1-V(s1),2)*pow(1-V(s3),1)*pow(1-V(s5),2)+2*V(a15)*(1)*pow(1-V(s1),2)*pow(1-V(s3),1)*pow(1-V(s4),2)}
.func fs4() = {2*V(a1)*(-1)*pow(1-V(s5),2)*pow(1+V(s4),1)*pow(1-V(s3),2)+2*V(a6)*(-1)*pow(1+V(s4),1)*pow(1+V(s1),2)*pow(1-V(s3),2)+2*V(a7)*(-1)*pow(1-V(s5),2)*pow(1+V(s1),2)*pow(1+V(s4),1)+2*V(a8)*(1)*pow(1-V(s5),2)*pow(1-V(s3),2)*pow(1-V(s4),1)+2*V(a9)*(-1)*pow(1+V(s5),2)*pow(1-V(s2),2)*pow(1+V(s4),1)+2*V(a11)*(1)*pow(1+V(s3),2)*pow(1-V(s5),2)*pow(1-V(s4),1)+2*V(a12)*(-1)*pow(1-V(s3),2)*pow(1+V(s4),1)*pow(1-V(s2),2)+2*V(a13)*(1)*pow(1+V(s2),2)*pow(1-V(s4),1)*pow(1-V(s5),2)+2*V(a15)*(1)*pow(1-V(s1),2)*pow(1-V(s3),2)*pow(1-V(s4),1)}
.func fs5() = {2*V(a1)*(1)*pow(1-V(s5),1)*pow(1+V(s4),2)*pow(1-V(s3),2)+2*V(a2)*(1)*pow(1-V(s5),1)*pow(1+V(s2),2)*pow(1+V(s3),2)+2*V(a3)*(1)*pow(1-V(s3),2)*pow(1-V(s5),1)*pow(1-V(s1),2)+2*V(a4)*(1)*pow(1+V(s2),2)*pow(1-V(s5),1)*pow(1-V(s3),2)+2*V(a5)*(-1)*pow(1+V(s2),2)*pow(1-V(s1),2)*pow(1+V(s5),1)+2*V(a7)*(1)*pow(1-V(s5),1)*pow(1+V(s1),2)*pow(1+V(s4),2)+2*V(a8)*(1)*pow(1-V(s5),1)*pow(1-V(s3),2)*pow(1-V(s4),2)+2*V(a9)*(-1)*pow(1+V(s5),1)*pow(1-V(s2),2)*pow(1+V(s4),2)+2*V(a10)*(-1)*pow(1+V(s3),2)*pow(1+V(s5),1)*pow(1+V(s2),2)+2*V(a11)*(1)*pow(1+V(s3),2)*pow(1-V(s5),1)*pow(1-V(s4),2)+2*V(a13)*(1)*pow(1+V(s2),2)*pow(1-V(s4),2)*pow(1-V(s5),1)+2*V(a14)*(1)*pow(1-V(s1),2)*pow(1-V(s3),2)*pow(1-V(s5),1)}

.func fa1() = {V(a1)*pow((1-V(s5))*(1+V(s4))*(1-V(s3)),2)}
.func fa2() = {V(a2)*pow((1-V(s5))*(1+V(s2))*(1+V(s3)),2)}
.func fa3() = {V(a3)*pow((1-V(s3))*(1-V(s5))*(1-V(s1)),2)}
.func fa4() = {V(a4)*pow((1+V(s2))*(1-V(s5))*(1-V(s3)),2)}
.func fa5() = {V(a5)*pow((1+V(s2))*(1-V(s1))*(1+V(s5)),2)}
.func fa6() = {V(a6)*pow((1+V(s4))*(1+V(s1))*(1-V(s3)),2)}
.func fa7() = {V(a7)*pow((1-V(s5))*(1+V(s1))*(1+V(s4)),2)}
.func fa8() = {V(a8)*pow((1-V(s5))*(1-V(s3))*(1-V(s4)),2)}
.func fa9() = {V(a9)*pow((1+V(s5))*(1-V(s2))*(1+V(s4)),2)}
.func fa10() = {V(a10)*pow((1+V(s3))*(1+V(s5))*(1+V(s2)),2)}
.func fa11() = {V(a11)*pow((1+V(s3))*(1-V(s5))*(1-V(s4)),2)}
.func fa12() = {V(a12)*pow((1-V(s3))*(1+V(s4))*(1-V(s2)),2)}
.func fa13() = {V(a13)*pow((1+V(s2))*(1-V(s4))*(1-V(s5)),2)}
.func fa14() = {V(a14)*pow((1-V(s1))*(1-V(s3))*(1-V(s5)),2)}
.func fa15() = {V(a15)*pow((1-V(s1))*(1-V(s3))*(1-V(s4)),2)}


.tran 0 300.000000 1u uic
.probe V(contra) V(contrd) V(s1) V(s2) V(s3) V(s4) V(s5)
\end{lstlisting}

\begin{lstlisting}[caption={LTspice netlist  for the digital memcomputing algorithm for the problem in Listing~\ref{list:1}}, label=list:3]
* parameters
.param alpha=5.000000 beta=20.000000 gamma=0.250000 delta=0.050000 epsilon=0.001000 xi=0.010000 xlmax=910000

* Control circuit
ESAT1 contra 0 value={fsat1()}
RSAT1 contra 0 100meg
ESAT2 contrd 0 value={fsat2()}
RSAT2 contrd 0 100meg

* Main variables
Cv1 v1 0 1 IC={-1+mc(1,1)}
Gv1 0 v1 value={fv1()*(u(1-V(v1))*u(fv1())+u(V(v1)+1)*u(-fv1()))}
Rv1 v1 0 100meg
Cv2 v2 0 1 IC={-1+mc(1,1)}
Gv2 0 v2 value={fv2()*(u(1-V(v2))*u(fv2())+u(V(v2)+1)*u(-fv2()))}
Rv2 v2 0 100meg
Cv3 v3 0 1 IC={-1+mc(1,1)}
Gv3 0 v3 value={fv3()*(u(1-V(v3))*u(fv3())+u(V(v3)+1)*u(-fv3()))}
Rv3 v3 0 100meg
Cv4 v4 0 1 IC={-1+mc(1,1)}
Gv4 0 v4 value={fv4()*(u(1-V(v4))*u(fv4())+u(V(v4)+1)*u(-fv4()))}
Rv4 v4 0 100meg
Cv5 v5 0 1 IC={-1+mc(1,1)}
Gv5 0 v5 value={fv5()*(u(1-V(v5))*u(fv5())+u(V(v5)+1)*u(-fv5()))}
Rv5 v5 0 100meg

* Short memory variables
Cs1 xs1 0 1 IC={0.5}
Gs1 0 xs1 value={fs1()*(u(1-V(xs1))*u(fs1())+u(V(xs1))*u(-fs1()))}
Rs1 xs1 0 100meg
Cs2 xs2 0 1 IC={0.5}
Gs2 0 xs2 value={fs2()*(u(1-V(xs2))*u(fs2())+u(V(xs2))*u(-fs2()))}
Rs2 xs2 0 100meg
Cs3 xs3 0 1 IC={0.5}
Gs3 0 xs3 value={fs3()*(u(1-V(xs3))*u(fs3())+u(V(xs3))*u(-fs3()))}
Rs3 xs3 0 100meg
Cs4 xs4 0 1 IC={0.5}
Gs4 0 xs4 value={fs4()*(u(1-V(xs4))*u(fs4())+u(V(xs4))*u(-fs4()))}
Rs4 xs4 0 100meg
Cs5 xs5 0 1 IC={0.5}
Gs5 0 xs5 value={fs5()*(u(1-V(xs5))*u(fs5())+u(V(xs5))*u(-fs5()))}
Rs5 xs5 0 100meg
Cs6 xs6 0 1 IC={0.5}
Gs6 0 xs6 value={fs6()*(u(1-V(xs6))*u(fs6())+u(V(xs6))*u(-fs6()))}
Rs6 xs6 0 100meg
Cs7 xs7 0 1 IC={0.5}
Gs7 0 xs7 value={fs7()*(u(1-V(xs7))*u(fs7())+u(V(xs7))*u(-fs7()))}
Rs7 xs7 0 100meg
Cs8 xs8 0 1 IC={0.5}
Gs8 0 xs8 value={fs8()*(u(1-V(xs8))*u(fs8())+u(V(xs8))*u(-fs8()))}
Rs8 xs8 0 100meg
Cs9 xs9 0 1 IC={0.5}
Gs9 0 xs9 value={fs9()*(u(1-V(xs9))*u(fs9())+u(V(xs9))*u(-fs9()))}
Rs9 xs9 0 100meg
Cs10 xs10 0 1 IC={0.5}
Gs10 0 xs10 value={fs10()*(u(1-V(xs10))*u(fs10())+u(V(xs10))*u(-fs10()))}
Rs10 xs10 0 100meg
Cs11 xs11 0 1 IC={0.5}
Gs11 0 xs11 value={fs11()*(u(1-V(xs11))*u(fs11())+u(V(xs11))*u(-fs11()))}
Rs11 xs11 0 100meg
Cs12 xs12 0 1 IC={0.5}
Gs12 0 xs12 value={fs12()*(u(1-V(xs12))*u(fs12())+u(V(xs12))*u(-fs12()))}
Rs12 xs12 0 100meg
Cs13 xs13 0 1 IC={0.5}
Gs13 0 xs13 value={fs13()*(u(1-V(xs13))*u(fs13())+u(V(xs13))*u(-fs13()))}
Rs13 xs13 0 100meg
Cs14 xs14 0 1 IC={0.5}
Gs14 0 xs14 value={fs14()*(u(1-V(xs14))*u(fs14())+u(V(xs14))*u(-fs14()))}
Rs14 xs14 0 100meg
Cs15 xs15 0 1 IC={0.5}
Gs15 0 xs15 value={fs15()*(u(1-V(xs15))*u(fs15())+u(V(xs15))*u(-fs15()))}
Rs15 xs15 0 100meg

* Long memory variables
Cl1 xl1 0 1 IC={1}
Gl1 0 xl1 value={fl1()*(u(xlmax-V(xl1))*u(fl1())+u(V(xl1)-1)*u(-fl1()))}
Rl1 xl1 0 100meg
Cl2 xl2 0 1 IC={1}
Gl2 0 xl2 value={fl2()*(u(xlmax-V(xl2))*u(fl2())+u(V(xl2)-1)*u(-fl2()))}
Rl2 xl2 0 100meg
Cl3 xl3 0 1 IC={1}
Gl3 0 xl3 value={fl3()*(u(xlmax-V(xl3))*u(fl3())+u(V(xl3)-1)*u(-fl3()))}
Rl3 xl3 0 100meg
Cl4 xl4 0 1 IC={1}
Gl4 0 xl4 value={fl4()*(u(xlmax-V(xl4))*u(fl4())+u(V(xl4)-1)*u(-fl4()))}
Rl4 xl4 0 100meg
Cl5 xl5 0 1 IC={1}
Gl5 0 xl5 value={fl5()*(u(xlmax-V(xl5))*u(fl5())+u(V(xl5)-1)*u(-fl5()))}
Rl5 xl5 0 100meg
Cl6 xl6 0 1 IC={1}
Gl6 0 xl6 value={fl6()*(u(xlmax-V(xl6))*u(fl6())+u(V(xl6)-1)*u(-fl6()))}
Rl6 xl6 0 100meg
Cl7 xl7 0 1 IC={1}
Gl7 0 xl7 value={fl7()*(u(xlmax-V(xl7))*u(fl7())+u(V(xl7)-1)*u(-fl7()))}
Rl7 xl7 0 100meg
Cl8 xl8 0 1 IC={1}
Gl8 0 xl8 value={fl8()*(u(xlmax-V(xl8))*u(fl8())+u(V(xl8)-1)*u(-fl8()))}
Rl8 xl8 0 100meg
Cl9 xl9 0 1 IC={1}
Gl9 0 xl9 value={fl9()*(u(xlmax-V(xl9))*u(fl9())+u(V(xl9)-1)*u(-fl9()))}
Rl9 xl9 0 100meg
Cl10 xl10 0 1 IC={1}
Gl10 0 xl10 value={fl10()*(u(xlmax-V(xl10))*u(fl10())+u(V(xl10)-1)*u(-fl10()))}
Rl10 xl10 0 100meg
Cl11 xl11 0 1 IC={1}
Gl11 0 xl11 value={fl11()*(u(xlmax-V(xl11))*u(fl11())+u(V(xl11)-1)*u(-fl11()))}
Rl11 xl11 0 100meg
Cl12 xl12 0 1 IC={1}
Gl12 0 xl12 value={fl12()*(u(xlmax-V(xl12))*u(fl12())+u(V(xl12)-1)*u(-fl12()))}
Rl12 xl12 0 100meg
Cl13 xl13 0 1 IC={1}
Gl13 0 xl13 value={fl13()*(u(xlmax-V(xl13))*u(fl13())+u(V(xl13)-1)*u(-fl13()))}
Rl13 xl13 0 100meg
Cl14 xl14 0 1 IC={1}
Gl14 0 xl14 value={fl14()*(u(xlmax-V(xl14))*u(fl14())+u(V(xl14)-1)*u(-fl14()))}
Rl14 xl14 0 100meg
Cl15 xl15 0 1 IC={1}
Gl15 0 xl15 value={fl15()*(u(xlmax-V(xl15))*u(fl15())+u(V(xl15)-1)*u(-fl15()))}
Rl15 xl15 0 100meg

* functions
.func Cm(x,y,z)={0.5*min(1-x,min(1-y,1-z))}
.func Cm1(x,y,z)={min(1-u(x),min(1-u(y),1-u(z)))}

.func fsat1()=Cm(V(v5),-V(v4),V(v3))+Cm(V(v5),-V(v2),-V(v3))+Cm(V(v3),V(v5),V(v1))+Cm(-V(v2),V(v5),V(v3))+Cm(-V(v2),V(v1),-V(v5))+Cm(-V(v4),-V(v1),V(v3))+Cm(V(v5),-V(v1),-V(v4))+Cm(V(v5),V(v3),V(v4))+Cm(-V(v5),V(v2),-V(v4))+Cm(-V(v3),-V(v5),-V(v2))+Cm(-V(v3),V(v5),V(v4))+Cm(V(v3),-V(v4),V(v2))+Cm(-V(v2),V(v4),V(v5))+Cm(V(v1),V(v3),V(v5))+Cm(V(v1),V(v3),V(v4))

.func fsat2()=Cm1(V(v5),-V(v4),V(v3))+Cm1(V(v5),-V(v2),-V(v3))+Cm1(V(v3),V(v5),V(v1))+Cm1(-V(v2),V(v5),V(v3))+Cm1(-V(v2),V(v1),-V(v5))+Cm1(-V(v4),-V(v1),V(v3))+Cm1(V(v5),-V(v1),-V(v4))+Cm1(V(v5),V(v3),V(v4))+Cm1(-V(v5),V(v2),-V(v4))+Cm1(-V(v3),-V(v5),-V(v2))+Cm1(-V(v3),V(v5),V(v4))+Cm1(V(v3),-V(v4),V(v2))+Cm1(-V(v2),V(v4),V(v5))+Cm1(V(v1),V(v3),V(v5))+Cm1(V(v1),V(v3),V(v4))

.func fv1() = {V(xl3)*V(xs3)*0.5*min(1-V(v3),1-V(v5))+(1+xi*V(xl3))*(1-V(xs3))*0.5*(1-V(v1))*if(V(v1) > V(v3), if(V(v1) > V(v5),1,0),0)+V(xl5)*V(xs5)*0.5*min(1+V(v2),1+V(v5))+(1+xi*V(xl5))*(1-V(xs5))*0.5*(1-V(v1))*if(V(v1) > -V(v2), if(V(v1) > -V(v5),1,0),0)+V(xl6)*V(xs6)*(-0.5)*min(1+V(v4),1-V(v3))+(1+xi*V(xl6))*(1-V(xs6))*0.5*(-1-V(v1))*if(-V(v1) > -V(v4), if(-V(v1) > V(v3),1,0),0)+V(xl7)*V(xs7)*(-0.5)*min(1-V(v5),1+V(v4))+(1+xi*V(xl7))*(1-V(xs7))*0.5*(-1-V(v1))*if(-V(v1) > V(v5), if(-V(v1) > -V(v4),1,0),0)+V(xl14)*V(xs14)*0.5*min(1-V(v3),1-V(v5))+(1+xi*V(xl14))*(1-V(xs14))*0.5*(1-V(v1))*if(V(v1) > V(v3), if(V(v1) > V(v5),1,0),0)+V(xl15)*V(xs15)*0.5*min(1-V(v3),1-V(v4))+(1+xi*V(xl15))*(1-V(xs15))*0.5*(1-V(v1))*if(V(v1) > V(v3), if(V(v1) > V(v4),1,0),0)}
.func fv2() = {V(xl2)*V(xs2)*(-0.5)*min(1-V(v5),1+V(v3))+(1+xi*V(xl2))*(1-V(xs2))*0.5*(-1-V(v2))*if(-V(v2) > V(v5), if(-V(v2) > -V(v3),1,0),0)+V(xl4)*V(xs4)*(-0.5)*min(1-V(v5),1-V(v3))+(1+xi*V(xl4))*(1-V(xs4))*0.5*(-1-V(v2))*if(-V(v2) > V(v5), if(-V(v2) > V(v3),1,0),0)+V(xl5)*V(xs5)*(-0.5)*min(1-V(v1),1+V(v5))+(1+xi*V(xl5))*(1-V(xs5))*0.5*(-1-V(v2))*if(-V(v2) > V(v1), if(-V(v2) > -V(v5),1,0),0)+V(xl9)*V(xs9)*0.5*min(1+V(v5),1+V(v4))+(1+xi*V(xl9))*(1-V(xs9))*0.5*(1-V(v2))*if(V(v2) > -V(v5), if(V(v2) > -V(v4),1,0),0)+V(xl10)*V(xs10)*(-0.5)*min(1+V(v3),1+V(v5))+(1+xi*V(xl10))*(1-V(xs10))*0.5*(-1-V(v2))*if(-V(v2) > -V(v3), if(-V(v2) > -V(v5),1,0),0)+V(xl12)*V(xs12)*0.5*min(1-V(v3),1+V(v4))+(1+xi*V(xl12))*(1-V(xs12))*0.5*(1-V(v2))*if(V(v2) > V(v3), if(V(v2) > -V(v4),1,0),0)+V(xl13)*V(xs13)*(-0.5)*min(1-V(v4),1-V(v5))+(1+xi*V(xl13))*(1-V(xs13))*0.5*(-1-V(v2))*if(-V(v2) > V(v4), if(-V(v2) > V(v5),1,0),0)}
.func fv3() = {V(xl1)*V(xs1)*0.5*min(1-V(v5),1+V(v4))+(1+xi*V(xl1))*(1-V(xs1))*0.5*(1-V(v3))*if(V(v3) > V(v5), if(V(v3) > -V(v4),1,0),0)+V(xl2)*V(xs2)*(-0.5)*min(1-V(v5),1+V(v2))+(1+xi*V(xl2))*(1-V(xs2))*0.5*(-1-V(v3))*if(-V(v3) > V(v5), if(-V(v3) > -V(v2),1,0),0)+V(xl3)*V(xs3)*0.5*min(1-V(v5),1-V(v1))+(1+xi*V(xl3))*(1-V(xs3))*0.5*(1-V(v3))*if(V(v3) > V(v5), if(V(v3) > V(v1),1,0),0)+V(xl4)*V(xs4)*0.5*min(1+V(v2),1-V(v5))+(1+xi*V(xl4))*(1-V(xs4))*0.5*(1-V(v3))*if(V(v3) > -V(v2), if(V(v3) > V(v5),1,0),0)+V(xl6)*V(xs6)*0.5*min(1+V(v4),1+V(v1))+(1+xi*V(xl6))*(1-V(xs6))*0.5*(1-V(v3))*if(V(v3) > -V(v4), if(V(v3) > -V(v1),1,0),0)+V(xl8)*V(xs8)*0.5*min(1-V(v5),1-V(v4))+(1+xi*V(xl8))*(1-V(xs8))*0.5*(1-V(v3))*if(V(v3) > V(v5), if(V(v3) > V(v4),1,0),0)+V(xl10)*V(xs10)*(-0.5)*min(1+V(v5),1+V(v2))+(1+xi*V(xl10))*(1-V(xs10))*0.5*(-1-V(v3))*if(-V(v3) > -V(v5), if(-V(v3) > -V(v2),1,0),0)+V(xl11)*V(xs11)*(-0.5)*min(1-V(v5),1-V(v4))+(1+xi*V(xl11))*(1-V(xs11))*0.5*(-1-V(v3))*if(-V(v3) > V(v5), if(-V(v3) > V(v4),1,0),0)+V(xl12)*V(xs12)*0.5*min(1+V(v4),1-V(v2))+(1+xi*V(xl12))*(1-V(xs12))*0.5*(1-V(v3))*if(V(v3) > -V(v4), if(V(v3) > V(v2),1,0),0)+V(xl14)*V(xs14)*0.5*min(1-V(v1),1-V(v5))+(1+xi*V(xl14))*(1-V(xs14))*0.5*(1-V(v3))*if(V(v3) > V(v1), if(V(v3) > V(v5),1,0),0)+V(xl15)*V(xs15)*0.5*min(1-V(v1),1-V(v4))+(1+xi*V(xl15))*(1-V(xs15))*0.5*(1-V(v3))*if(V(v3) > V(v1), if(V(v3) > V(v4),1,0),0)}
.func fv4() = {V(xl1)*V(xs1)*(-0.5)*min(1-V(v5),1-V(v3))+(1+xi*V(xl1))*(1-V(xs1))*0.5*(-1-V(v4))*if(-V(v4) > V(v5), if(-V(v4) > V(v3),1,0),0)+V(xl6)*V(xs6)*(-0.5)*min(1+V(v1),1-V(v3))+(1+xi*V(xl6))*(1-V(xs6))*0.5*(-1-V(v4))*if(-V(v4) > -V(v1), if(-V(v4) > V(v3),1,0),0)+V(xl7)*V(xs7)*(-0.5)*min(1-V(v5),1+V(v1))+(1+xi*V(xl7))*(1-V(xs7))*0.5*(-1-V(v4))*if(-V(v4) > V(v5), if(-V(v4) > -V(v1),1,0),0)+V(xl8)*V(xs8)*0.5*min(1-V(v5),1-V(v3))+(1+xi*V(xl8))*(1-V(xs8))*0.5*(1-V(v4))*if(V(v4) > V(v5), if(V(v4) > V(v3),1,0),0)+V(xl9)*V(xs9)*(-0.5)*min(1+V(v5),1-V(v2))+(1+xi*V(xl9))*(1-V(xs9))*0.5*(-1-V(v4))*if(-V(v4) > -V(v5), if(-V(v4) > V(v2),1,0),0)+V(xl11)*V(xs11)*0.5*min(1+V(v3),1-V(v5))+(1+xi*V(xl11))*(1-V(xs11))*0.5*(1-V(v4))*if(V(v4) > -V(v3), if(V(v4) > V(v5),1,0),0)+V(xl12)*V(xs12)*(-0.5)*min(1-V(v3),1-V(v2))+(1+xi*V(xl12))*(1-V(xs12))*0.5*(-1-V(v4))*if(-V(v4) > V(v3), if(-V(v4) > V(v2),1,0),0)+V(xl13)*V(xs13)*0.5*min(1+V(v2),1-V(v5))+(1+xi*V(xl13))*(1-V(xs13))*0.5*(1-V(v4))*if(V(v4) > -V(v2), if(V(v4) > V(v5),1,0),0)+V(xl15)*V(xs15)*0.5*min(1-V(v1),1-V(v3))+(1+xi*V(xl15))*(1-V(xs15))*0.5*(1-V(v4))*if(V(v4) > V(v1), if(V(v4) > V(v3),1,0),0)}
.func fv5() = {V(xl1)*V(xs1)*0.5*min(1+V(v4),1-V(v3))+(1+xi*V(xl1))*(1-V(xs1))*0.5*(1-V(v5))*if(V(v5) > -V(v4), if(V(v5) > V(v3),1,0),0)+V(xl2)*V(xs2)*0.5*min(1+V(v2),1+V(v3))+(1+xi*V(xl2))*(1-V(xs2))*0.5*(1-V(v5))*if(V(v5) > -V(v2), if(V(v5) > -V(v3),1,0),0)+V(xl3)*V(xs3)*0.5*min(1-V(v3),1-V(v1))+(1+xi*V(xl3))*(1-V(xs3))*0.5*(1-V(v5))*if(V(v5) > V(v3), if(V(v5) > V(v1),1,0),0)+V(xl4)*V(xs4)*0.5*min(1+V(v2),1-V(v3))+(1+xi*V(xl4))*(1-V(xs4))*0.5*(1-V(v5))*if(V(v5) > -V(v2), if(V(v5) > V(v3),1,0),0)+V(xl5)*V(xs5)*(-0.5)*min(1+V(v2),1-V(v1))+(1+xi*V(xl5))*(1-V(xs5))*0.5*(-1-V(v5))*if(-V(v5) > -V(v2), if(-V(v5) > V(v1),1,0),0)+V(xl7)*V(xs7)*0.5*min(1+V(v1),1+V(v4))+(1+xi*V(xl7))*(1-V(xs7))*0.5*(1-V(v5))*if(V(v5) > -V(v1), if(V(v5) > -V(v4),1,0),0)+V(xl8)*V(xs8)*0.5*min(1-V(v3),1-V(v4))+(1+xi*V(xl8))*(1-V(xs8))*0.5*(1-V(v5))*if(V(v5) > V(v3), if(V(v5) > V(v4),1,0),0)+V(xl9)*V(xs9)*(-0.5)*min(1-V(v2),1+V(v4))+(1+xi*V(xl9))*(1-V(xs9))*0.5*(-1-V(v5))*if(-V(v5) > V(v2), if(-V(v5) > -V(v4),1,0),0)+V(xl10)*V(xs10)*(-0.5)*min(1+V(v3),1+V(v2))+(1+xi*V(xl10))*(1-V(xs10))*0.5*(-1-V(v5))*if(-V(v5) > -V(v3), if(-V(v5) > -V(v2),1,0),0)+V(xl11)*V(xs11)*0.5*min(1+V(v3),1-V(v4))+(1+xi*V(xl11))*(1-V(xs11))*0.5*(1-V(v5))*if(V(v5) > -V(v3), if(V(v5) > V(v4),1,0),0)+V(xl13)*V(xs13)*0.5*min(1+V(v2),1-V(v4))+(1+xi*V(xl13))*(1-V(xs13))*0.5*(1-V(v5))*if(V(v5) > -V(v2), if(V(v5) > V(v4),1,0),0)+V(xl14)*V(xs14)*0.5*min(1-V(v1),1-V(v3))+(1+xi*V(xl14))*(1-V(xs14))*0.5*(1-V(v5))*if(V(v5) > V(v1), if(V(v5) > V(v3),1,0),0)}

.func fs1() = {beta*(V(xs1)+epsilon)*(Cm(V(v5),-V(v4),V(v3))-gamma)}
.func fs2() = {beta*(V(xs2)+epsilon)*(Cm(V(v5),-V(v2),-V(v3))-gamma)}
.func fs3() = {beta*(V(xs3)+epsilon)*(Cm(V(v3),V(v5),V(v1))-gamma)}
.func fs4() = {beta*(V(xs4)+epsilon)*(Cm(-V(v2),V(v5),V(v3))-gamma)}
.func fs5() = {beta*(V(xs5)+epsilon)*(Cm(-V(v2),V(v1),-V(v5))-gamma)}
.func fs6() = {beta*(V(xs6)+epsilon)*(Cm(-V(v4),-V(v1),V(v3))-gamma)}
.func fs7() = {beta*(V(xs7)+epsilon)*(Cm(V(v5),-V(v1),-V(v4))-gamma)}
.func fs8() = {beta*(V(xs8)+epsilon)*(Cm(V(v5),V(v3),V(v4))-gamma)}
.func fs9() = {beta*(V(xs9)+epsilon)*(Cm(-V(v5),V(v2),-V(v4))-gamma)}
.func fs10() = {beta*(V(xs10)+epsilon)*(Cm(-V(v3),-V(v5),-V(v2))-gamma)}
.func fs11() = {beta*(V(xs11)+epsilon)*(Cm(-V(v3),V(v5),V(v4))-gamma)}
.func fs12() = {beta*(V(xs12)+epsilon)*(Cm(V(v3),-V(v4),V(v2))-gamma)}
.func fs13() = {beta*(V(xs13)+epsilon)*(Cm(-V(v2),V(v4),V(v5))-gamma)}
.func fs14() = {beta*(V(xs14)+epsilon)*(Cm(V(v1),V(v3),V(v5))-gamma)}
.func fs15() = {beta*(V(xs15)+epsilon)*(Cm(V(v1),V(v3),V(v4))-gamma)}

.func fl1() = {alpha*(Cm(V(v5),-V(v4),V(v3))-delta)}
.func fl2() = {alpha*(Cm(V(v5),-V(v2),-V(v3))-delta)}
.func fl3() = {alpha*(Cm(V(v3),V(v5),V(v1))-delta)}
.func fl4() = {alpha*(Cm(-V(v2),V(v5),V(v3))-delta)}
.func fl5() = {alpha*(Cm(-V(v2),V(v1),-V(v5))-delta)}
.func fl6() = {alpha*(Cm(-V(v4),-V(v1),V(v3))-delta)}
.func fl7() = {alpha*(Cm(V(v5),-V(v1),-V(v4))-delta)}
.func fl8() = {alpha*(Cm(V(v5),V(v3),V(v4))-delta)}
.func fl9() = {alpha*(Cm(-V(v5),V(v2),-V(v4))-delta)}
.func fl10() = {alpha*(Cm(-V(v3),-V(v5),-V(v2))-delta)}
.func fl11() = {alpha*(Cm(-V(v3),V(v5),V(v4))-delta)}
.func fl12() = {alpha*(Cm(V(v3),-V(v4),V(v2))-delta)}
.func fl13() = {alpha*(Cm(-V(v2),V(v4),V(v5))-delta)}
.func fl14() = {alpha*(Cm(V(v1),V(v3),V(v5))-delta)}
.func fl15() = {alpha*(Cm(V(v1),V(v3),V(v4))-delta)}

.tran 0 300.000000 1u uic
.probe V(contra) V(contrd) V(v1) V(v2) V(v3) V(v4) V(v5)
\end{lstlisting}

\end{document}